\documentclass[journal]{IEEEtran}
\usepackage{color}
\usepackage{cite}
\usepackage{url}

\ifCLASSINFOpdf
\else
\fi
\usepackage{caption}
\usepackage{amsfonts,amssymb}
\usepackage{graphicx}
\usepackage{epstopdf}
\usepackage{mathrsfs}
\usepackage{amsmath}
\usepackage{latexsym}
\usepackage{booktabs}
\usepackage{float}
\usepackage{subfigure}
\usepackage{algorithm, algorithmic}
\usepackage{bbm}
\usepackage[amsmath,thmmarks]{ntheorem}
\usepackage{bm}
\DeclareCaptionFont{blue}{\color{blue}}

\begin{document}
%
\title{\LARGE Performance Optimization for Semantic Communications: An Attention-based Reinforcement Learning Approach}




%
\author{\normalsize{{Yining Wang,} \emph{Student Member, IEEE},  {Mingzhe Chen,} \emph{Member,~IEEE}, \\
{Tao Luo,} \emph{Senior Member,~IEEE}, {Walid Saad,} \emph{Fellow,~IEEE}, {Dusit Niyato,} \emph{Fellow,~IEEE}\\
{H. Vincent Poor,} \emph{Life Fellow,~IEEE}, and {Shuguang Cui,} \emph{Fellow,~IEEE}}\vspace*{-3em}\\
\thanks{Y. Wang and T. Luo are with the Beijing Laboratory of Advanced Information Network, Beijing University of Posts and Telecommunications, Beijing, 100876, China (e-mail: \protect\url{wyy0206@bupt.edu.cn}; \protect\url{tluo@bupt.edu.cn}).}
\thanks{M. Chen and H. V. Poor are with the Department of Electrical and Computer Engineering, Princeton University, Princeton, NJ, 08544, USA (e-mail: \protect\url{mingzhec@princeton.edu}; \protect\url{poor@princeton.edu}).}
\thanks{W. Saad is with the Wireless@VT, Bradley Department of Electrical and Computer Engineering, Virginia Tech, Arlington, VA, 22203, USA (e-mail: \protect\url{walids@vt.edu}).}
\thanks{D. Niyato is with the School of Computer Science and Engineering (SCSE), NTU, Singapore (e-mail: \protect\url{dniyato@ntu.edu.sg}).}
\thanks{S. Cui is currently with the School of Science and Engineering (SSE), the Future Network of Intelligence Institute (FNii), and the Guangdong Provincial Key Laboratory of Future Networks of Intelligence, the Chinese University of Hong Kong, and Shenzhen Research Institute of Big Data, Shenzhen, China, 518172; he is also affiliated with Peng Cheng Laboratory, Shenzhen, China, 518066 (e-mail: \protect\url{shuguangcui@cuhk.edu.cn}).}
\thanks{This work was supported in part by the National Natural Science Foundation of China under Grant 62171047, in part by the National Research Foundation, Singapore and Infocomm Media Development Authority under its Future Communications Research \& Development Programme, in part by the AI Singapore Programme (AISG) (AISG2-RP-2020-019), in part by Singapore Ministry of Education (MOE) Tier 1 (RG16/20), in part by the National Key R\&D Program of China with grant No. 2018YFB1800800, in part by the Basic Research Project No. HZQB-KCZYZ-2021067 of Hetao Shenzhen-HK S\&T Cooperation Zone, in part by Shenzhen Outstanding Talents Training Fund~202002, in part by Guangdong Research Projects No. 2017ZT07X152 and No. 2019CX01X104, and in part by the Guangdong Provincial Key Laboratory of Future Networks of Intelligence (Grant No. 2022B1212010001).}
\thanks{A preliminary version of this work \cite{wyy2021GC} is published in the Proceedings of the 2021 IEEE International Global Communications Conference (GLOBECOM).}
}


\maketitle

\begin{abstract}
In this paper, a semantic communication framework is proposed for textual data transmission.
In the studied model, a base station (BS) extracts the semantic information from textual data, and transmits it to each user. 
The semantic information is modeled by a knowledge graph (KG) that consists of a set of semantic triples.
After receiving the semantic information, each user recovers the original text using a graph-to-text generation model.
To measure the performance of the considered semantic communication framework, a metric of semantic similarity (MSS) that jointly captures the semantic accuracy and completeness of the recovered text is proposed.
Due to wireless resource limitations, the BS may not be able to transmit the entire semantic information to each user and satisfy the transmission delay constraint. 
Hence, the BS must select an appropriate resource block for each user as well as determine and transmit part of the semantic information to the users. 
As such, we formulate an optimization problem whose goal is to maximize the total MSS by jointly optimizing the resource allocation policy and determining the partial semantic information to be transmitted.
To solve this problem, a proximal-policy-optimization-based reinforcement learning (RL) algorithm integrated with an attention network is proposed.
The proposed algorithm can evaluate the importance of each triple in the semantic information using an attention network and then, build a relationship between the importance distribution of the triples in the semantic information and the total MSS.
Compared to traditional RL algorithms, the proposed algorithm can dynamically adjust its learning rate thus ensuring convergence to a locally optimal solution.
Simulation results show that the proposed framework can reduce by 41.3\% data that the BS needs to transmit and improve by two-fold the total MSS compared to a standard communication network without using semantic communication techniques.
\end{abstract}


%
\IEEEpeerreviewmaketitle

\vspace{-0.2cm}
\section{Introduction}
New wireless applications such as extended reality and digital twinning are generating unprecedented amounts of data (at zetta-bytes scale) which will strain the capacity of wireless networks \cite{6G_walid}.
To support these human-centered services and applications, wireless networks must be carefully designed based on content, human-related requirements, human-related knowledge, and experience-based metrics~\cite{Semantic_6G}.
One of the solutions for these challenges is \emph{semantic communication}, which allows the meaning of the data (behind digital bits) to be extracted and exploited during wireless transmission \cite{Semantic_6G, Deep_Semantic, Semantic_Magizine}.
Semantic communication has recently attracted significant interest due to its advantages in terms of providing human-oriented services and improving communication efficiency.
However, deploying semantic communications in wireless networks faces several challenges including extraction of data meaning, semantic-oriented resource allocation, semantic encoding and decoding, measurement of semantic communication performance, and security of semantic communications.

\vspace{-0.2cm}
\subsection{Related Works}
Recently, a number of works such as  \cite{Deep_Semantic, Semantic_IOT, Semantic_Significance, Semantic_AoI1, Semantic_Hierarchical, Semantic_Magizine} investigated the semantic communications over wireless networks. In particular, in \cite{Deep_Semantic}, the authors  proposed a deep learning approach to optimize the mutual information between the original and decoded signals.
The authors in \cite{Semantic_IOT} designed a distributed semantic communication system for capacity-limited networks.
In \cite{Deep_Semantic} and \cite{Semantic_IOT}, the authors used a deep neural network based autoencoder to encode data and considered the output of the autoencoder as the semantic information of the data. 
However, the output of the autoencoder being a vector of decimals is incomprehensible for humans and does not have any physical meaning.
In \cite{Semantic_Significance} and \cite{Semantic_AoI1}, the authors developed a semantic communication framework for real-time control systems and considered the control signals as the semantic information of the data.
However, using control signals as semantic information is not applicable for text or image data transmission since control signals cannot represent the content related to a text or image.
The authors in \cite{Semantic_Hierarchical} used a hierarchical structure to represent the relationships among the objects in the real world so as to understand the meaning underlying the data to be transmitted. 
In~\cite{Semantic_Magizine}, the authors provided an overview on the use of semantic detection and knowledge modeling techniques for semantic information extraction.
However, both works in \cite{Semantic_Magizine} and \cite{Semantic_Hierarchical} ignored the design of semantic metrics that can measure the performance of semantic communications.

In \cite{NSW_2011_theory, Semantic_AoI2, Semantic_Speech, Semantic_Game}, the authors proposed a number of metrics for evaluating the performance of semantic communication systems.
The work in \cite{NSW_2011_theory} derived the semantic channel capacity based on the mutual information between the transmitted symbols and the desired meaning.
In \cite{Semantic_AoI2}, the authors proposed a semantic communication system whose goal is to minimize the average age of incorrect information (AoII) that captures the accuracy and freshness of the real-time data.
The authors in \cite{Semantic_Speech} used the perceptual evaluation of speech distortion as well as speech signal metrics to evaluate the performance of a semantic-based speech transmission framework.
The work in \cite{Semantic_Game} defined the semantic distance between two words based on the distribution of words in a given corpus so as to measure the semantic similarity of two sentences. 
While interesting, the metrics developed in \cite{NSW_2011_theory, Semantic_AoI2, Semantic_Speech, Semantic_Game} only capture whether the information in the received data is correct without guaranteeing that the received data contains all the information in the original data to be transmitted.
Meanwhile, these prior works \cite{NSW_2011_theory, Semantic_AoI2, Semantic_Speech, Semantic_Game} assumed that all the semantic information extracted from the original data can be transmitted over the network and did not consider the semantic communication over wireless resource-constrained networks. 
In practice, due to wireless resource limitations (e.g., bandwidth), the BS may not be able to transmit the entire semantic information to all users \cite{resource}.
A number of existing works such as in \cite{VR_TWC_chenmingzhe_huan, resource_zhanghaijun, JSTSP_WSH, resource_liuyuanwei, resource_xiongkangdusit} studied the problem of resource allocation in wireless networks. 
However, the solutions in these existing works \cite{VR_TWC_chenmingzhe_huan, resource_zhanghaijun, JSTSP_WSH, resource_liuyuanwei, resource_xiongkangdusit} cannot be applied in semantic communications since they do not consider the effects of semantic information extraction and transmission on resource allocation.


\vspace{-0.2cm}
\subsection{Contributions}
The main contribution of this work is a novel semantic communication framework that jointly considers semantic information extraction and transmission, data recovery, as well as performance evaluation.
The key contributions are summarized as follows:
\begin{itemize}
\item We consider a semantic communication network in which a base station (BS) uses semantic communication techniques to extract the meaning of the text data and transmits it to its associated users.
The meaning of the text data is defined as the semantic information and modeled by a knowledge graph (KG) that consists of a set of semantic triples in the form of ``\emph{entity-relation-entity}".
Based on the received semantic information, each user uses a graph-to-text generation model to recover the original text.
\item 
We propose a mathematical metric, called the metric of semantic similarity (MSS), to capture the semantic communication performance. 
The proposed MSS can measure whether the information in the recovered text is correct and whether the recovered text contains all the information in the original text.
\item To satisfy the delay constraint, the BS must optimize the resource allocation for each user and transmit a part of the semantic information to each user.
This problem is formulated as an optimization problem whose goal is to maximize the total MSS by optimizing the resource allocation policy and determining the semantic information that needs to be transmitted.
\item To solve this problem, we propose an attention proximal policy optimization (APPO) algorithm that can evaluate the importance of each triple in the semantic information.
Then, the proposed algorithm can analyze the relationship between the importance distribution of the triples in the semantic information and the total MSS, thus finding the effective policies for resource allocation and semantic information transmission.
\item We analyze the convergence of the proposed APPO algorithm. The analytical result shows that the APPO algorithm is guaranteed to converge to a locally optimal solution of the studied total MSS maximization problem.
\end{itemize} 
Simulation results show that, compared to a standard communication network that does not consider semantic communications, the proposed APPO algorithm can reduce the number of words that the BS needs to transmit by up to 41.3\% while achieving a 2-fold improvement in the total MSS.
To our knowledge, \emph{this is the first work that introduces a mathematical model for semantic communication enabled wireless networks and optimizes resource allocation and semantic information transmission to improve the performance of semantic-driven wireless networks}.

The rest of this paper is organized as follows. 
The proposed semantic communication framework and the problem formulation are described in Section \ref{sec:2}. 
The use of APPO algorithm for resource allocation and partial semantic information determination is introduced in Section \ref{sec:3}. 
In Section \ref{sec:4}, the numerical results are presented and discussed. 
Finally, conclusions are drawn in Section \ref{sec:5}.

\vspace{-0.2cm}
\section{System Model and Problem Formulation}
\label{sec:2}

\begin{table}\footnotesize
\newcommand{\tabincell}[2]{\begin{tabular}{@{}#1@{}}#1.6\end{tabular}}
\renewcommand\arraystretch{1}
\caption[table]{{List of notation}}
\centering
\begin{tabular}{|c|c|c|c|c|c|}
\hline
\!\textbf{Notation}\! \!\!& \textbf{Description}\\
\hline
$U $ & Number of users \\
\hline
 $Q$ &  Number of downlink orthogonal RBs\\
\hline
$L_i$ & Original text needed to transmit to user $i$ \\
\hline
${\bm{\alpha}}_i$ & RB allocation vector for user $i$ \\
\hline
$w_{i,n}$ & Token $n$ in text $L_i$\\
\hline
$c_i({\bm{\alpha}}_i)$& Downlink channel capacity \\
\hline
$N_i$ & Number of tokens in text $L_i$\\
\hline
$W$ & Bandwidth of each RB \\
\hline
$e_{i,j}$ & Entity $j$ in text $L_i$\\
\hline
$P$ & Transmit power of the BS \\
\hline
$E_i$ & Number of entities in text $L_i$\\
\hline
$I_q$ & Interference of RB $q$ \\
\hline
$r_{i,jk}$ & Relation between entity $e_{i,j}$ and entity $e_{i,k}$\\
\hline
$N_0$ & Noise power spectral density \\
\hline
$R_i$ & Number of relations in text $L_i$\\
\hline
$T$ & Transmission delay threshold \\
\hline 
${\cal{G}}_i$ & Semantic information of text $L_i$\\ 
\hline 
$\phi_i$ & Channel gain between the BS and user $i$ \\
\hline 
$\bm{\varepsilon}_i^g$ & Semantic triple $g$ in ${\cal{G}}_i$\\
\hline 
${L}'_i({\bm{\alpha}}_i,{\cal{G}}'_i)$ & Text recovered by user $i$ \\
\hline 
$Z({\cal{G}}_i)$ & Number of tokens in ${\cal{G}}_i$\\
\hline 
$M_i$ & Number of tokens in text ${L}'_i({\bm{\alpha}}_i,{\cal{G}}'_i)$ \\
\hline 
${\cal{G}}'_i$ & Partial semantic information transmitted to user $i$\\
\hline 
$O$ & Number of bits used to represent a token \\
\hline 
$A_i({\bm{\alpha}}_i,{\cal{G}}'_i)$ & Semantic accuracy\\
\hline
$E_i({\bm{\alpha}}_i,{\cal{G}}'_i)$ & Metric of semantic similarity \\
\hline 
$R_i({\bm{\alpha}}_i,{\cal{G}}'_i)$ & Semantic completeness\\
\hline 
$\xi_i$ & Penalty for short text\\
\hline 
$\varphi$ & Parameter adjusting $A_i({\bm{\alpha}}_i,{\cal{G}}'_i)$ and $R_i({\bm{\alpha}}_i,{\cal{G}}'_i)$\\
\hline 
\end{tabular}
\end{table}

\begin{figure}[t]
\centering
\setlength{\belowcaptionskip}{-0.5cm}
\includegraphics[width=8cm]{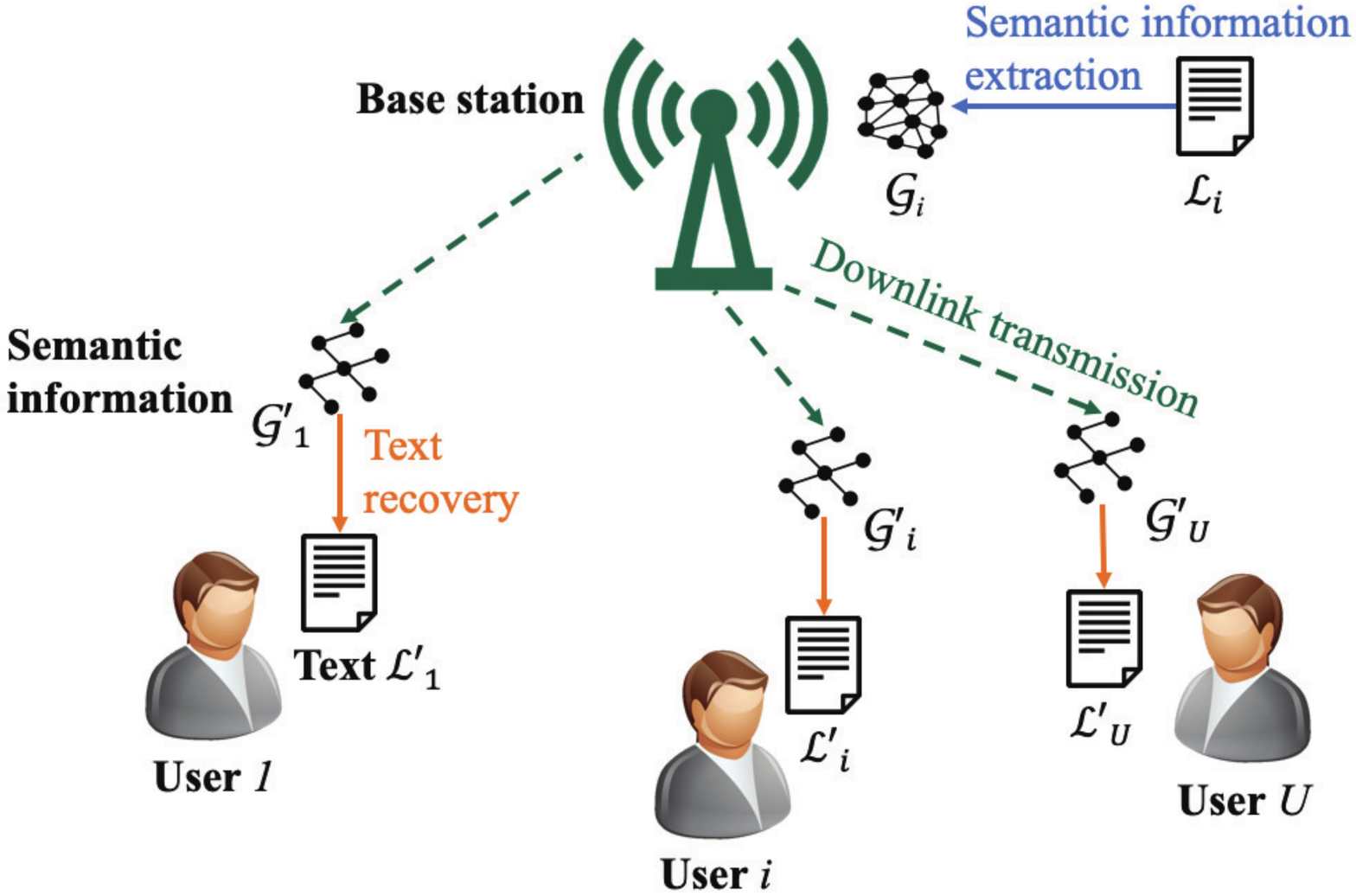}
\caption{Illustration of the proposed semantic communication framework.}
\label{fig1}
\end{figure}

We consider a wireless network that consists of a BS using semantic communication techniques to transmit the meaning of text data to a set $\cal{U}$ of $U$ users. 
Hereinafter, the meaning of the text data transmitted over wireless links is referred to \emph{semantic information}.  
To achieve semantic communications, the BS extracts the semantic information from the original text and sends it to the corresponding user\footnote{In this work, we only consider the textual data transmission. 
One can easily extend the proposed model to other types of data such as audio data and image data \cite{goal_Walid, speech1, image1}.}.
Then, each user recovers the text based on the received semantic information.
In particular, the considered semantic communication process consists of four phases (shown in Fig.~\ref{fig1}): a) semantic information extraction, b) semantic-oriented resource allocation and semantic information selection, c) original text data recovery, and d) semantic similarity evaluation.
Next, the process of semantic communications is first introduced.
Then, a semantic similarity model is proposed to measure the performance of semantic communications.

\vspace{-0.2cm}
\subsection{Semantic Information Extraction}
A \emph{token} $w_{i,n}$ is used to represent a word, a symbol, or a punctuation in the text data. 
As a result, the text data that the BS needs to transmit to user $i$ consists of a sequence of tokens, as follows: 
\begin{equation}
\label{text_t}
{L}_i=\{w_{i,1},w_{i,2},\ldots,w_{i,n},\ldots,w_{i,N_i}\}, \forall w_{i,n} \in \cal{V},
\end{equation}
where $\cal{V}$ is the set of tokens in a corpus and $N_i$ is the number of tokens in ${L}_i$.
For example, if the BS requirs to transmit ``\emph{Little girls are playing.}" to user $i$, then, we have ${L}_i=\{[{\rm{little}}], [{\rm{girls}}],[{\rm{are}}],[{\rm{playing}}], [.]\}$, where $w_{i,1}=[{\rm{little}}]$, $w_{i,2}=[{\rm{girls}}]$, $w_{i,3}=[{\rm{are}}]$, $w_{i,4}=[{\rm{playing}}]$, and $w_{i,5}=[.]$.

\begin{figure*}[t]
\centering
\setlength{\belowcaptionskip}{-0.5cm}
\includegraphics[width=13cm]{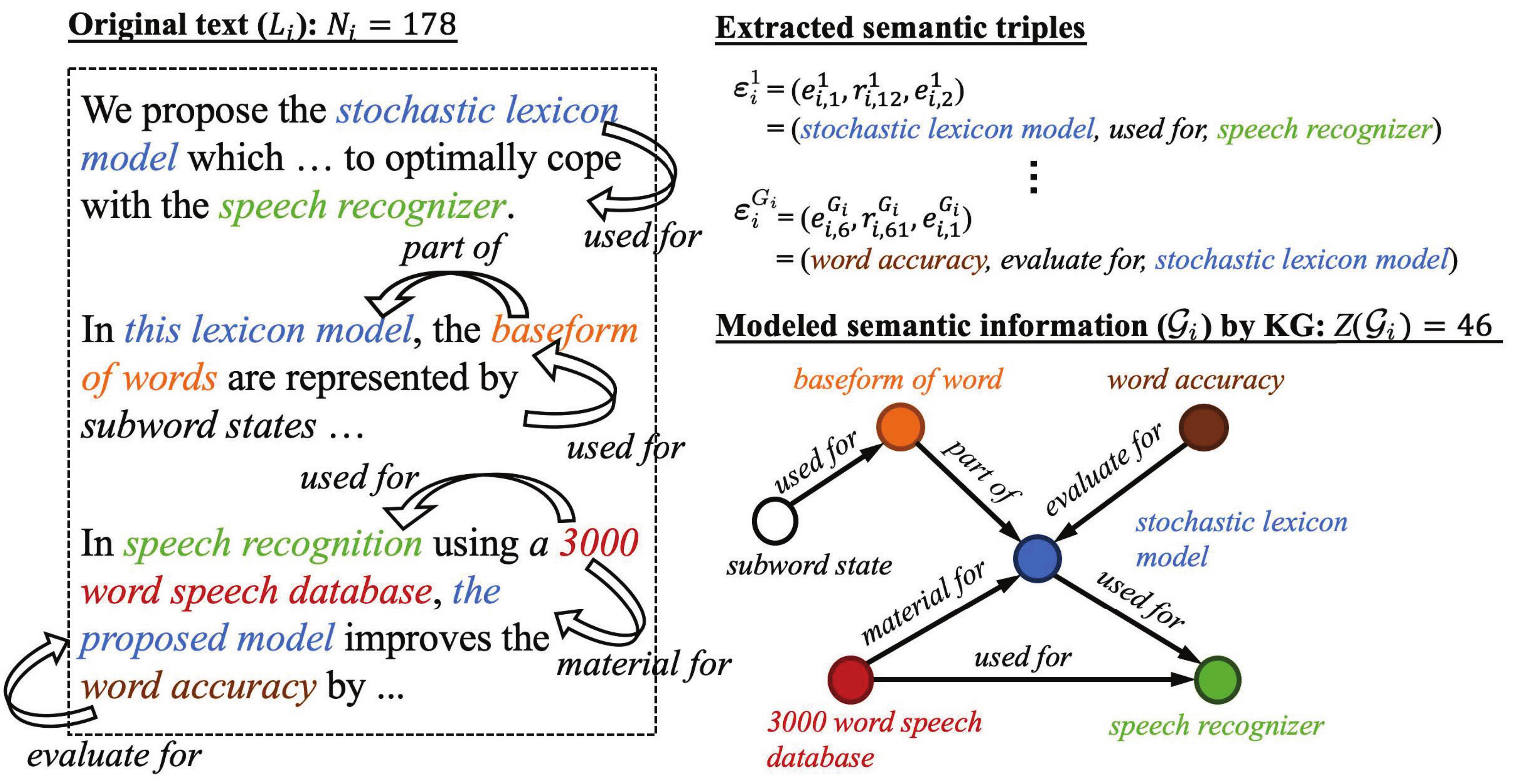}
\caption{An example of an original text and the extracted semantic information.}
\label{fig2}
\end{figure*}

In our model, the semantic information extracted from a text data is modeled by a KG \cite{KG}.
Hence, the semantic information consists of a set of nodes and a set of edges, as shown in Fig.~\ref{fig2}. 
In particular,  each node in the semantic information is an \emph{entity} that refers to an object or a concept in the real world.
Hereinafter, we define entity $j$ that consists of a subsequence of text $L_i$ as $e_{i,j}$.
For example, in Fig.~\ref{fig2}, ``\emph{baseform of word}" is an entity that consists of four tokens in the example text.
An information extraction system such as the scientific information extractor in~\cite{IE_system} can be used to recognize the set ${\cal{E}}_i$ of $E_i$ entities in text ${L}_i$.

Edges are the \emph{relations} between each pair of entities.
Given a pair of recognized entities $(e_{i,j},e_{i,k}), j \ne k$, the BS must find the relation $r_{i,jk} \in {\cal{R}}_i$ between them, where ${\cal{R}}_i$ is the set of $R_i$ relations involved in text $L_i$.
For example, in Fig.~\ref{fig2}, the relation between entity ``\emph{baseform of word}" and ``\emph{stochastic lexicon model}" can be formulated as ``\emph{part of}".
Note that the relations (i.e., the edges of the semantic information) are directional and, hence, we have $r_{i,jk} \ne r_{i,kj}$.
We assume that there is a predefined set $\cal{R}$ containing all relations in the texts and each relation is a two-token sequence such as ``\emph{part of}" and ``\emph{evaluate for}", as done in \cite{IE_system}.
Hence, given a pair of entities $e_{i,j}$ and $e_{i,k}$ in the original text $L_i$, the relation $r_{i,jk}$ between $e_{i,j}$ and $e_{i,k}$ can be obtained by classification algorithms such as convolutional neural networks~\cite{CNN_relation}.
Here, the input of the relation classification algorithm is the sentence containing the pair of entities in the original text and the output is the predefined two-token relation that can summarize the description of the entities in the input sentence.
We assume that the information extraction system can recognize all the entities in the original text.
Hence, the extracted triples in the form of ``\emph{entity-relation-entity}" can represent \emph{the meaning} of the original text.

Based on the recognized entities and the extracted relations, the semantic information of text ${L}_i$ can be modeled as
\begin{equation}
\label{KG_t}
{\cal{G}}_i\!=\!\{\bm{\varepsilon}_i^1, \ldots, \bm{\varepsilon}_i^g,\ldots, \bm{\varepsilon}_i^{G_i}\},  
\end{equation}
where $\bm{\varepsilon}_i^g=(e_{i,j}^g,r_{i,jk}^g,e_{i,k}^g),\forall e_{i,j}^g,e_{i,k}^g \!\in\! {\cal{E}}_i, j \!\ne\! k, \forall r_{i,jk}^g \!\in\! {\cal{R}}_i$ is a semantic triple and $G_i$ is the number of semantic triples in ${\cal{G}}_i$.
Since each entity (e.g., $e_{i,j}^g$ or $e_{i,k}^g$) and each relation $r_{i,jk}^g$ consist of a sequence of tokens (e.g., $e_{i,j}\!=\!\{[{\rm{baseform}}],[{\rm{of}}],[{\rm{word}}]\}$, $e_{i,k}\!=\!\{[{\rm{stochastic}}],[{\rm{lexicon}}],[{\rm{model}}]\}$, and $r_{i,jk}\!=\!\{[{\rm{part}}],[{\rm{of}}]\}$), triple $\bm{\varepsilon}_i^g$ can be expressed as $\bm{\varepsilon}_i^g=\{v_{i,1}^g,\ldots,v_{i,b}^g,\ldots,v_{i,B_i^g}^g\}$, where $v_{i,b}^g \in {\cal{V}}$ is token $b$ in semantic triple $\bm{\varepsilon}_i^g$ and $B_i^g={S_{i,j}^g\!+\!S_{i,k}^g\!+\!{S_{i,jk}^g}}$ with $S_{i,j}^g$ and $S_{i,jk}^g$ being the number of tokens in entity $e_{i,j}^g$ and relation $r_{i,jk}^g$, respectively.
Then, the number of tokens in semantic information ${\cal{G}}_i$ is
\begin{equation}
\label{num_token_t}
Z({\cal{G}}_i)=\sum\limits_{g=1}^{G_i} \left({{S_{i,j}^g} + {S_{i,k}^g} + {S_{i,jk}^g}}\right).
\end{equation}
From Fig.~\ref{fig2}, we see that the data size of the extracted semantic information is much smaller than the data size of the original text (i.e., $Z({\cal{G}}_i) \ll N_i$).
The reason is that the two-token relations between the entity pairs in ${\cal{G}}_i$ can reduce the redundant context in original text $L_i$.

\vspace{-0.2cm}
\subsection{Transmission Model}
An orthogonal frequency division multiple access (OFDMA) technique is used for semantic information transmission.
In our model, a set $\cal{Q}$ of $Q$ downlink orthogonal resource blocks (RBs) can be allocated to serve the users. 
The RB allocation vector of user $i$ is ${\bm{\alpha}}_i=[\alpha_{i,1},\ldots,\alpha_{i,q},\ldots,\alpha_{i,Q}]$, where $\alpha_{i,q} \in \{0,1\}$.
Here, $\alpha_{i,q}=1$ implies that RB $q$ is allocated to user $i$; otherwise, we have $\alpha_{i,q}=0$.
In our model, we assume that each RB can only be allocated to one user and each user can only occupy one RB.
Then, we have
\begin{equation}
\label{allocation}
\sum\limits_{q = 1}^Q {\alpha_{i,q}} \leqslant 1, \forall i \in {\cal{U}}; \;\;\;\;\sum\limits_{i = 1}^U {\alpha_{i,q}} \leqslant 1, \forall q \in {\cal{Q}}.
\end{equation}
The downlink channel capacity of the BS transmitting semantic information ${\cal{G}}_i$ to user $i$ is given as \cite{chen_2021}
\begin{equation}
\label{capacity}
c_i({\bm{\alpha}}_i)=\sum\limits_{q = 1}^Q{\alpha_{i,q}W{\log_2}\left(1 + \frac{{P}{\phi_i}}{I_q+WN_0}\right)}
\end{equation}
where $W$ is the bandwidth of each RB, $P$ is the transmit power of the BS, $I_q$ represents the interference caused by BSs that are located in other service areas and use RB $q$, $N_0$ is the noise power spectral density, and $\phi_i={\gamma_i}{{d_i}^{-2}}$ is the channel gain between the BS and user $i$ with $\gamma_i$ being the Rayleigh fading parameters and $d_i$ being the distance between the BS and user $i$.
Here, we assume that the transmit power $P$ of each user $i$ is a constant and one can easily consider the optimization of transmit power by extending the studied model.
We also assume that the transmission delay between the BS and each user $i$ is limited to $D$. 
Given the data rate $c_i({\bm{\alpha}}_i)$, the maximum number of tokens that can be transmitted within the transmission delay $D$ is determined.
Hence, to satisfy the transmission delay constraint, the BS must determine partial semantic information (i.e., a subset of triples) to be transmitted.

\vspace{-0.2cm}
\subsection{Text Recovery}
The partial semantic information that the BS transmits to user $i$ is given as
\begin{equation}
\label{KG_r}
{\cal{G}}'_i\!=\!\{{\bm{\varepsilon}'}_i^1, \ldots, {\bm{\varepsilon}'}_i^h,\ldots, {\bm{\varepsilon}'}_i^{H_i}\} \subset {\cal{G}}_i, 
\end{equation}
where ${\bm{\varepsilon}'}_i^h\!=\!({e'}_{i,j}^h, {r'}_{i,jk}^h, {e'}_{i,k}^h)$ and $H_i$ is the number of selected semantic triples in ${\cal{G}}'_i$.
Given the transmission delay threshold $D$, the selected semantic information ${\cal{G}}'_i$ should satisfy the delay constraint as follows: 
\begin{equation}
\label{data_size}
\frac{Z({\cal{G}}'_i)O}{c_i({\bm{\alpha}}_i)} \leqslant D,
\end{equation}
where $Z({\cal{G}}'_i)\!=\!\sum\limits_{h=1}^{H_i} \left({{S_{i,j}^h} + {S_{i,k}^h} + 2}\right)$ and $O$ is the number of bits used to represent each token.
After each user $i$ receives the partial semantic information ${\cal{G}}'_i$, a graph-to-text generation model, such as the graph transformer in \cite{graph_transformer}, can be used to recover the coherent multi-sentence text from ${\cal{G}}'_i$.
We assume that the graph-to-text generation model \cite{graph_transformer} is well-trained and shared among all users.
The text recovered by user $i$ based on ${\cal{G}}'_i$ is
\begin{equation}
\label{text_r}
{L}'_i({\bm{\alpha}}_i,{\cal{G}}'_i)=\{w'_{i,0},w'_{i,1},\ldots,w'_{i,m},\ldots,w'_{i,M_i}\}, 
\end{equation}
where $M_i$ is the number of tokens in the recovered text ${L}'_i$.

\vspace{-0.2cm}
\subsection{Semantic Similarity Model}
To measure the quality of semantic communications, we propose a metric of semantic similarity (MSS).
Compared to the existing cosine similarity metric \cite{Deep_Semantic}, the proposed MSS can capture semantic similarity while avoiding semantic errors caused by word vectorization.
Hereinafter, we define the degree of the information in the recovered text being correct as \emph{semantic accuracy} and the degree of the recovered text containing the information of the original text as \emph{semantic completeness}.
Compared to the existing bilingual evaluation understudy (BLEU) metric \cite{BLEU} that only captures the semantic accuracy of the recovered text, the proposed MSS  jointly captures the semantic accuracy and completeness of the recovered text.
A method based on token matching is introduced to calculate the semantic accuracy and completeness \cite{METEOR}.
The semantic accuracy of recovered text ${L}'_i({\bm{\alpha}}_i,{\cal{G}}'_i)$ is defined as
\begin{equation}
\label{precision}
A_i({\bm{\alpha}}_i,{\cal{G}}'_i)=\frac{\sum\limits_{m = 1}^{M_i}{\min \left(\sigma\left({L}'_i({\bm{\alpha}}_i,{\cal{G}}'_i),w'_{i,m}\right),\sigma({L}_i,w{'_{i,m}})\right)}}{\sum\limits_{m = 1}^{M_i} {\sigma\left({L}'_i({\bm{\alpha}}_i,{\cal{G}}'_i),w'_{i,m}\right)}},
\end{equation}
where $\sigma({{L}'_i({\bm{\alpha}}_i,{\cal{G}}'_i)}_i,w'_{i,m})$ and $\sigma({L}_i,w{'_{i,m}})$ is the number of occurrences of token $w'_{i,m}$ in recovered text ${L}'_i({\bm{\alpha}}_i,{\cal{G}}'_i)$ and in original text $L_i$, respectively, and
$\min \!\left(\sigma\!\left({L}'_i({\bm{\alpha}}_i,{\cal{G}}'_i),w'_{i,m}\right)\!,\!\sigma({L}_i,w{'_{i,m}})\right)$ indicates the number of correct occurrences of token $w'_{i,m}$ in recovered text ${L}'_i({\bm{\alpha}}_i,{\cal{G}}'_i)$. 
For example, if token $w'_{i,m}=[\rm{you}]$ occurs twice in the recovered text and once in the original text, only one of the two occurrences of $w'_{i,m}=[\rm{you}]$ is correct in the recovered text. 
In~(\ref{precision}), $\sum\limits_{m = 1}^{M_i}{\min \left(\sigma\left({L}'_i({\bm{\alpha}}_i,{\cal{G}}'_i),w'_{i,m}\right),\sigma({L}_i,w{'_{i,m}})\right)}$ represents the sum of the number of correct occurrences of each token in the recovered text and $\sum\limits_{m = 1}^{M_i} {\sigma\left({L}'_i({\bm{\alpha}}_i,{\cal{G}}'_i),w'_{i,m}\right)}$ represents the sum of the number of occurrences of each token in the recovered text.
Hence, $A_i({\bm{\alpha}}_i,{\cal{G}}'_i)$ is defined as the ratio of the sum of the number of correct occurrences of each token to the sum of the number of occurrences of each token.

The semantic completeness of the recovered text is defined as 
\begin{equation}
\label{recall}
R_i({\bm{\alpha}}_i,{\cal{G}}'_i)=\frac{\sum\limits_{m = 1}^{M_i}{\min \left(\sigma\left({L}'_i({\bm{\alpha}}_i,{\cal{G}}'_i),w'_{i,m}\right),\sigma({L}_i,w{'_{i,m}})\right)}}{\sum\limits_{n = 1}^{N_i} {\sigma(L_i,w_{i,n})}}.
\end{equation}
In (\ref{recall}), $\sum\limits_{n = 1}^{N_i} {\sigma(L_i,w_{i,n})}$ represents the the sum of the number of occurrences of each token in the original text.
Hence, $R_i({\bm{\alpha}}_i,{\cal{G}}'_i)$ is defined as the ratio of the sum of the number of correct occurrences of each token in the recovered text to the sum of the number of occurrences of each token in the original text.
Next, we use an example to explain the differences between the semantic accuracy and the semantic completeness more clearly.
For example, $L_i=\{[{\rm{little}}],[{\rm{girls}}], [{\rm{are}}],[{\rm{playing}}]\}$.
For a recovered text ${L}'_i({\bm{\alpha}}_i,{\cal{G}}'_i)=\left\{[{\rm{girls}}],[{\rm{are}}], [{\rm{playing}}]\right\}$, we have $A_i({\bm{\alpha}}_i,{\cal{G}}'_i)=\frac{1+1+1}{1+1+1}=1$ and $R_i({\bm{\alpha}}_i,{\cal{G}}'_i)=\frac{1+1+1}{1+1+1+1}=\frac{3}{4}$, respectively. 

Based on (\ref{precision}) and (\ref{recall}), the MSS of recovered text ${L}'_i({\bm{\alpha}}_i,{\cal{G}}'_i)$ can be given as
\begin{equation}
\label{MSS}
E_i({\bm{\alpha}}_i,{\cal{G}}'_i)=\xi_i \frac{A_i({\bm{\alpha}}_i,{\cal{G}}'_i)R_i({\bm{\alpha}}_i,{\cal{G}}'_i)}{\varphi A_i({\bm{\alpha}}_i,{\cal{G}}'_i)+(1-\varphi)R_i({\bm{\alpha}}_i,{\cal{G}}'_i)},
\end{equation}
where $\varphi \in (0,1)$ is a weight parameter used to adjust the contributions of semantic accuracy and completeness to the MSS.
$\xi_i$ is an additional penalty for short text and can be represented by \cite{BLEU} 
\begin{equation}
\label{penalty}
\xi_i = \left\{ {\begin{array}{*{20}{l}}
{\;\;\;\;1,\;\;\;\;M_i \geqslant N_i,}\\
{e^{1-\frac{N_i}{M_i}}, \;M_i < N_i.}
\end{array}} \right.
\end{equation}
Changing the value of $\xi_i$, each user can efficiently use of the received partial semantic information for text recovery.
From (\ref{MSS}), we see that the proposed MSS used to evaluate the recovered text can control the tradeoff between the semantic accuracy and the semantic completeness. 
In particular, for a given token $w{'_{i,m}}$ in the recovered text ${L}'_i({\bm{\alpha}}_i,{\cal{G}}'_i)$, if it appears more times in ${L}'_i({\bm{\alpha}}_i,{\cal{G}}'_i)$ than $L_i$, then $A_i({\bm{\alpha}}_i,{\cal{G}}'_i)$ decreases; otherwise, $R_i({\bm{\alpha}}_i,{\cal{G}}'_i)$ decreases.

\vspace{-0.2cm}
\subsection{Problem Formulation}
Given the defined system model, our goal is to maximize the total MSS of all the texts recovered by the users while satisfying the transmission delay requirement.
This maximization problem includes the RB allocation optimization and the partial semantic information determination.
The MSS maximization problem is formulated as follows:
\begin{subequations}
\label{optimization}
\begin{align}\tag{13}
&{\mathop {\max }\limits_{{\bm{\alpha}}_i,{\cal{G}}'_i} \;\;\sum\limits_{i=1}^U{E_i({\bm{\alpha}}_i,{\cal{G}}'_i)},}\\
&{\;\;{\rm{s}}.{\rm{t}}.\;\;\alpha_{i,q}\in \{0,1\},\forall i \in {\cal U}, \forall q \in {\cal Q},}\\
&{\;\;\;\;\;\;\;\;\;\sum\limits_{q = 1}^Q {\alpha_{i,q}} \leqslant 1, \forall i \in {\cal{U}},}\\
&{\;\;\;\;\;\;\;\;\;\sum\limits_{i = 1}^U {\alpha_{i,q}} \leqslant 1, \forall q \in {\cal{Q}},}\\
&{\;\;\;\;\;\;\;\;\;\frac{Z_i({\cal{G}}'_i)O}{c_i({\bm{\alpha}}_i)} \leqslant D,\forall i \in {\cal U},}
\end{align}
\end{subequations}
where the constraints in (\ref{optimization}a), (\ref{optimization}b), and (\ref{optimization}c) guarantee that each user can only occupy one RB and each RB can only be allocated to one user for semantic information transmission.
The constraint in (\ref{optimization}d) is the delay requirement of semantic information transmission. 
Problem (\ref{optimization}) is challenging to solve by the traditional algorithms such as greedy algorithms due to the following reasons. 
First, from (\ref{MSS}), we see that the objective function of problem (\ref{optimization}) depends not only on the transmitted partial semantic information ${\cal{G}}'_i$ and the RB allocation ${\bm{\alpha}}_i$, but also on the text generation model which is implemented by neural networks. 
Second, the relationship between the non-convex objective function (i.e., the total MSS) and the optimization variables (i.e., ${\cal{G}}'_i$ and ${\bm{\alpha}}_i$) cannot be characterized accurately.
Third, the coupling between the semantic information selection and the RB allocation further complicates the MSS maximization problem.
In particular, the size of the selected partial semantic information for each user depends on the allocated RB.
Meanwhile, the RB allocation depends on the importance of different semantic information in terms of the MSS improvement.
To solve (\ref{optimization}), we propose an attention network based reinforcement learning (RL) algorithm that enables the BS to evaluate the importance of semantic triples and optimize the RB allocation and the partial semantic information selection based on the importance of triples so as to improve the total MSS of all recovered texts.


\vspace{-0.2cm}
\section{Attention RL for Semantic Information Selection and Resource Allocation}
\label{sec:3}
We now introduce a proximal policy optimization (PPO)-based RL algorithm \cite{PPO} integrated with an attention network \cite{attention}, called attention proximal policy optimization (APPO).
First, for each semantic triple $\bm{\varepsilon}_i^g$, we use an attention network to calculate the corresponding importance value. 
Since the triples that are highly correlated with the original text are important for text recovery, the importance of a semantic triple is defined as the correlation between the triple and the original text.
Hereinafter, we use an importance vector $\bm{f}_i({\cal{G}}_i)$ to represent the \emph{importance distribution} of the triples in each semantic information ${\cal{G}}_i$.
Based on the importance evaluation, the proposed APPO algorithm enables the BS to analyze the relationship between the importance distribution $\bm{f}_i({\cal{G}}_i)$ and the total MSS so as to optimize the \emph{policy} for the RB allocation and semantic information selection.
Here, since the trained text recovery model can be considered as the prior knowledge of the texts to transmitted, it is shared by the users and the BS.
Hence, the BS can obtain the total MSS once the RB allocation and the partial semantic information to transmit are determined.
We first introduce the use of the attention networks to calculate the importance of the triples in each semantic information.
Then, we explain the components of the APPO algorithm and the process of using our proposed APPO algorithm to optimize the RB allocation for each user and determine the partial semantic information to be transmitted.
Finally, we show the complexity and convergence of the proposed APPO algorithm.

\vspace{-0.4cm}
\subsection{Attention Network for Importance Evaluation}
\label{sec:3.1}
Given semantic information ${\cal{G}}_i$ that consists of $G_i$ triples, an attention network is used to capture the importance of each triple $\bm{\varepsilon}_i^g \in {\cal{G}}_i$.
Then, the importance distribution $\bm{f}_i({\cal{G}}_i)$ of semantic information ${\cal{G}}_i$ can be obtained by normalizing the values of importance.
In particular, an attention network consists of an input layer, a hidden layer, an output layer, and a softmax layer, as shown in Fig.~\ref{attention}.
Next, we introduce each layer of an attention network.
\begin{itemize}
\item \emph{Input layer}: To obtain the importance of semantic triple $\bm{\varepsilon}_i^g$, the BS first needs to vectorize each token $v_{i,b}^g$ in triple $\bm{\varepsilon}_i^g$ and each token $w_{i,n}$ in original text $L_i$ as done in \cite{BERT}. 
We define the vector used to represent token $v_{i,b}^g$ as $\bm{x}_{i,b}^g \in \mathbb{R}^{D_x}$ and the vector used to represent token $w_{i,n}$ as $\bm{x}_{i,n} \in \mathbb{R}^{D_x}$, where $D_x$ is the dimension of each token vector.
\begin{figure}[t]
\centering
\setlength{\belowcaptionskip}{-0.5cm}
\includegraphics[width=8cm]{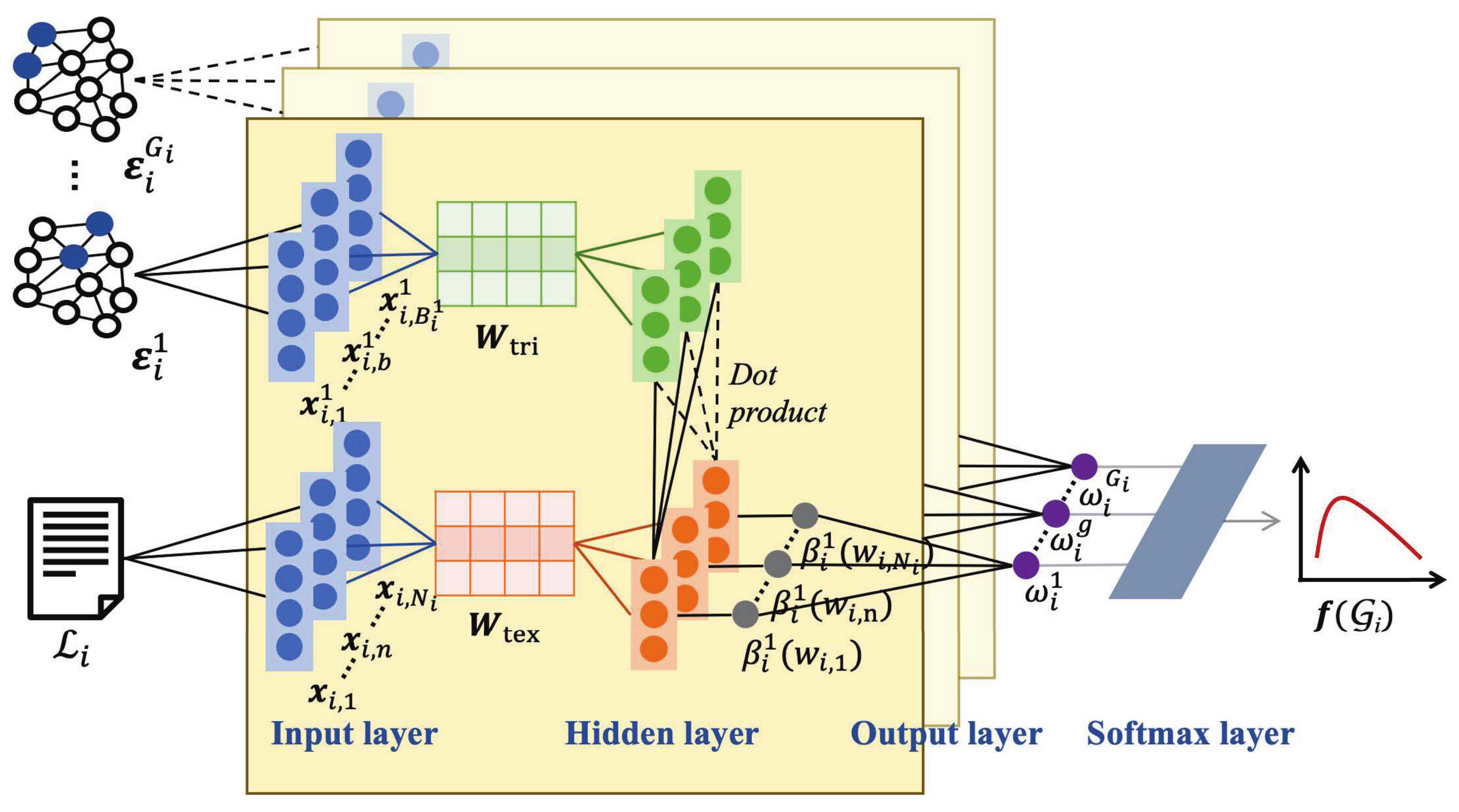}
\caption{The architecture of the attention networks.}
\label{attention}
\end{figure}
The input of an attention network is a sequence of token vectors $\bm{X}_{i}^g=(\bm{x}_{i,1}^g, \ldots, \bm{x}_{i,b}^g, \ldots, \bm{x}_{i,B_i^g}^g)$ that represent semantic triple $\bm{\varepsilon}_i^g$ and a sequence of token vectors $\bm{X}_{i}^L=(\bm{x}_{i,1}, \ldots,\bm{x}_{i,n}, \ldots,\bm{x}_{i,N_i})$ that represent original text $L_i$.
\item \emph{Hidden layer}: The hidden layer is used to find the correlation between each token $v_{i,b}^g \in \bm{\varepsilon}_i^g$ and each token $w_{i,n}$ in original text $L_i$.
Given the token vectors, the correlation between $v_{i,b}^g$ and $w_{i,n}$ can be given as
\begin{equation}
\label{correlation_token}
\psi(\bm{x}_{i,b}^g,\bm{x}_{i,n})={(\bm{W}_{{\rm{tri}}}\bm{x}_{i,b}^g)^{\mathsf{T}}}(\bm{W}_{{\rm{tex}}}\bm{x}_{i,n}),
\end{equation}
where $\bm{W}_{{\rm{tri}}} \in \mathbb{R}^{D_a \times D_x}$ and $\bm{W}_{{\rm{tex}}} \in \mathbb{R}^{D_a \times D_x}$ are both parameter matrices of the attention network with $D_a \times D_x$ being the size of the parameter matrices. 
Then, the correlation between triple $\bm{\varepsilon}_i^g$ in ${\cal{G}}_i$ and token $w_{i,n}$ in $L_i$ can be given as
\begin{equation}
\label{correlation}
\beta_i^g(\bm{x}_{i,n})=\sum\limits_{b = 1}^{B_i^g}{\frac{\psi(\bm{x}_{i,b}^g,\bm{x}_{i,n})}{B_i^g}},
\end{equation}
where $B_i^g$ is the number of tokens in triple $\bm{\varepsilon}_i^g$. 
\item \emph{Output layer}: The output of an attention network is the importance of semantic triple $\bm{\varepsilon}_i^g$, as follows: 
\begin{equation}
\label{importance}
\varsigma_i^g=\sum\limits_{n = 1}^{N_i} \beta_i^g(\bm{x}_{i,n}).
\end{equation}
\item \emph{Softmax layer}: Since the importance $\omega_i^g$ obtained by an attention network can be any value, the BS cannot directly determine the resource allocation and the partial semantic information to be transmitted based on $\omega_i^g$.
A softmax layer is used to normalize each importance $\omega_i^g$ so as to obtain the importance distribution of the triples in semantic information ${\cal{G}}_i$ as follows
\begin{equation}
\label{distribution}
\bm{f}({\cal{G}}_i)=\frac{e^{\bm{\varsigma}_i}}{\sum\limits_{g = 1}^{G_i} e^{\varsigma_i^g}}=[y_i^1, \ldots,y_i^g,\ldots,y_i^{G_i}],
\end{equation}
where $e$ is the Euler number, $\bm{\varsigma}_i=[\varsigma_i^1, \ldots,\varsigma_i^g,\ldots,\varsigma_i^{G_i}]$. 
Here, note that $\bm{f}({\cal{G}}_i)$ is an importance vector whose length is the number of the semantic triples in ${\cal{G}}_i$.
From (\ref{distribution}), we see that each element $y_i^g \in [0,1]$ represents the normalized importance of the corresponding semantic triple $\bm{\varepsilon}_i^g$ and $\sum\limits_{g = 1}^{G_i} y_i^g=1$.
\end{itemize}
Here, we note that, given the importance distribution $\bm{f}({\cal{G}}_i)$, the BS still cannot use the traditional algorithms such as greedy algorithms to directly obtain the optimal RB allocation and determine the partial semantic information to be transmitted.
The reason is that the MSS of each user depends on the received semantic information and the partial semantic information that can be transmitted depends on the RB allocation.
The BS is not aware of the relationship between each importance distribution $\bm{f}({\cal{G}}_i)$ and the optimal RB allocation.

\vspace{-0.2cm}
\subsection{Components of the APPO Algorithm}
Next, we introduce the components of the proposed APPO algorithm. 
The APPO algorithm consists of five components: a) agent, b) actions, c) states, d) policy, and e) reward, which are specified as follows: 
\begin{itemize}
\item \emph{Agent}: Our agent is the BS that determines the RB allocation and the partial semantic information to be transmitted for each user.
\item \emph{Actions}: We define the action of the agent as a vector $\bm{a}=[\bm{\alpha}_1,\ldots,\bm{\alpha}_i, \ldots, \bm{\alpha}_U]$ that represents the RB allocation for all users.
The action space $\cal{A}$ is the set of all optional actions that satisfy the constraints in (\ref{optimization}).
Here, once the RB allocated for each user is determined, the BS can transmit the triples in the corresponding semantic information in the order of the importance until the delay threshold is reached.
Hence, given the RB allocation $\bm{\alpha}_i$, the partial semantic information ${\cal{G}}'_i$ to be transmitted to user $i$ consists of the most important triples in ${\cal{G}}_i$ while satisfying $\frac{Z_i({\cal{G}}'_i)O}{c_i({\bm{\alpha}}_i)} \leqslant D$.
For example, we consider ${\cal{G}}_i=\{$(``\emph{stochastic lexicon model, used for, speech recognizer}"), (``\emph{baseform of word, part of, stochastic lexicon model}")$\}$ and hence, $\bm{\varepsilon}_i^1=$ (``\emph{stochastic lexicon model, used for, speech recognizer}") and $\bm{\varepsilon}_i^2=$ (``\emph{baseform of word, part of, stochastic lexicon model}").
Based on (\ref{distribution}), we have ${\bm{f}}({{\cal{G}}_i})=[0.7,0.3]$. 
We also set that only 10 tokens can be transmitted over the wireless network given the determined RB allocation $\bm{\alpha}_i$ and the delay constraint $D$.
Then, the BS will transmit the most important triple $\bm{\varepsilon}_i^1$ to the user $i$.
Hence, each ${\cal{G}}'_i$ is determined based on each $\bm{\alpha}_i$ in the selected action $\bm{a}$.
\item \emph{States}: The state defined as $\bm{s}=[\bm{f}({\cal{G}}_1),\ldots, \bm{f}({\cal{G}}_U)]$ is the importance distribution of all semantic information.
The state space $\cal{S}$ is a continuous space whose size depends on the number of users and the number of triples in each semantic information.
\item \emph{Policy}: The policy is the probability of the agent choosing each action given state $\bm{s}$. 
The APPO algorithm uses a deep neural network (DNN) parameterized by $\bm{\theta}$ to build the relationship between the input state $\bm{s}$ and the output policy that can achieve the maximum total MSS.
Then, the policy can be expressed as ${\bm{\pi}}_{\bm{\theta}}(\bm{s},\bm{a})=P(\bm{a}|\bm{s})$. 
\item \emph{Reward}: The reward of choosing action $\bm{a}$ based on state $\bm{s}$ is $R(\bm{a}|\bm{s})=\sum\limits_{i = 1}^U {E_i({\bm{\alpha}}_i,{\cal{G}}'_i)}$ which is the total MSS resulting from action $\bm{a}$ at state $\bm{s}$.
Since the reward function of the proposed APPO algorithm is equivalent to the objective function of problem (\ref{optimization}) and the APPO algorithm aims to maximize the reward, the proposed APPO algorithm can solve the total MSS maximization problem (\ref{optimization}).
\end{itemize}

\vspace{-0.2cm}
\subsection{APPO for Total MSS Maximization}
Next, we introduce the entire training process for the proposed APPO algorithm that allows us to solve problem (\ref{optimization}).
The proposed APPO algorithm is trained offline, which means that the RB allocation and semantic information selection policy is trained using the historical data of semantic communications.
Different from the attention policy gradient (APG) algorithm \cite{wyy2021GC} that uses a static learning rate for policy update, the proposed APPO algorithm can dynamically adjust the learning rate according to the difference between the old policy and the updated policy.
Hence, the proposed APPO algorithm can improve the total MSS over the training process and is guaranteed to converge to a locally optimal solution of problem (\ref{optimization}).
In addition, different from the APG algorithm of \cite{wyy2021GC} that samples new training data based on the policy updated at each iteration, the proposed APPO enables the BS to perform the RB allocation sampling and the policy update asynchronously. 
In other words, the proposed APPO algorithm can use historical sampled actions and their rewards for model training thus reducing the overhead caused by continuously sampling new actions when the policy is iteratively updated.
In particular, the BS stores a historical policy ${\bm{\pi}}_{{\bm{\theta}}^*}(\bm{s},\bm{a})$ (i.e., the policy updated at a past iteration) and samples $K$ RB allocation vectors according to ${\bm{\pi}}_{{\bm{\theta}}^*}(\bm{s},\bm{a})$.
The set of collected actions is ${{\cal{K}}}=\{\bm{a}^*_{1},\ldots,\bm{a}^*_{k},\ldots,\bm{a}^*_{K}\}$ that is used to train the policy ${\bm{\pi}}_{{\bm{\theta}}}(\bm{s},\bm{a})$ for total MSS maximization.
The expected reward of policy ${\bm{\pi}}_{{\bm{\theta}}}(\bm{s},\bm{a})$ that the BS aims to optimize is defined as
\begin{equation}
\label{expected_reward}
\begin{aligned}
\bar{A}({\bm{\theta}})&={\mathbb{E}}_{\bm{a} \sim {\bm{\pi}}_{{\bm{\theta}}}(\bm{s},\bm{a})}\left({R(\bm{a}|\bm{s})}\right),\\
&=\int {{R(\bm{a}|\bm{s})}{\bm{\pi}}_{{\bm{\theta}}}(\bm{s},\bm{a})} d\bm{a},\\
&=\int {{R(\bm{a}|\bm{s})}\frac{{\bm{\pi}}_{{\bm{\theta}}}(\bm{s},\bm{a})}{{\bm{\pi}}_{{\bm{\theta}}^*}(\bm{s},\bm{a})}{\bm{\pi}}_{{\bm{\theta}}^*}(\bm{s},\bm{a})} d\bm{a},\\
&={\mathbb{E}}_{\bm{a} \sim {\bm{\pi}}_{{\bm{\theta}}^*}(\bm{s},\bm{a})}\left({R(\bm{a}|\bm{s})}\frac{{\bm{\pi}}_{{\bm{\theta}}}(\bm{s},\bm{a})}{{\bm{\pi}}_{{\bm{\theta}}^*}(\bm{s},\bm{a})}\right),\\
&\approx \frac{1}{K}\sum\limits_{k = 1}^K {{R(\bm{a}^*_k|\bm{s})}\frac{{\bm{\pi}}_{{\bm{\theta}}}(\bm{s},\bm{a}^*_k)}{{\bm{\pi}}_{{\bm{\theta}}^*}(\bm{s},\bm{a}^*_k)}},
\end{aligned}
\end{equation} 
where ${\mathbb{E}}_{\bm{a} \sim {\bm{\pi}}_{{\bm{\theta}}}(\bm{s},\bm{a})}(\cdot)$ is the expectation with respect to action $\bm{a}$ following the policy ${\bm{\pi}}_{{\bm{\theta}}}(\bm{s},\bm{a})$.
In~(\ref{expected_reward}), in order to ensure that the set of sampled actions ${{\cal{D}}}$ can be used to evaluate and iteratively update the policy ${\bm{\pi}}_{{\bm{\theta}}}(\bm{s},\bm{a})$, the difference between the stored policy ${\bm{\pi}}_{{\bm{\theta}}^*}(\bm{s},\bm{a})$ and the policy ${\bm{\pi}}_{{\bm{\theta}}}(\bm{s},\bm{a})$ to be updated must be controlled~\cite{TRPO}.
Hence, the goal of optimizing the policy ${\bm{\pi}}_{{\bm{\theta}}}(\bm{s},\bm{a})$ is to maximize the total MSS with the penalty that captures the difference between ${\bm{\pi}}_{{\bm{\theta}}^*}(\bm{s},\bm{a})$ and ${\bm{\pi}}_{{\bm{\theta}}}(\bm{s},\bm{a})$.
Therefore, we have
\begin{equation}
\label{loss}
\mathop {\max }\limits_{\bm{\theta}} J(\bm{\theta}),
\end{equation}
where $J(\bm{\theta})=\bar{A}(\bm{\theta})-\lambda f_{\rm{KL}}\left({\bm{\pi}}_{{\bm{\theta}}^*}(\bm{s},\bm{a}),{\bm{\pi}}_{{\bm{\theta}}}(\bm{s},\bm{a})\right)$ with $\lambda$ being the penalty coefficient and $f_{\rm{KL}}\left({\bm{\pi}}_{{\bm{\theta}}^{*}}(\bm{s},\bm{a}),{\bm{\pi}}_{{\bm{\theta}}}(\bm{s},\bm{a})\right)$ being the Kullback–Leibler divergence (KLD) that represents the difference between ${\bm{\pi}}_{{\bm{\theta}}^{*}}(\bm{s},\bm{a})$ and ${\bm{\pi}}_{{\bm{\theta}}}(\bm{s},\bm{a})$.
 At each iteration $t$, the policy ${\bm{\pi}}_{{\bm{\theta}}}(\bm{s},\bm{a})$ will be updated using the standard gradient ascent method so as to improve the total MSS.
 The policy update rule is given as
\begin{equation}
\label{SGA}
{\bm{\theta}}^{(t)} \leftarrow {\bm{\theta}}^{(t-1)}+\delta \nabla_{\bm{\theta}} J({\bm{\theta}}),
\end{equation}
where ${\bm{\theta}}^{(t)}$ is the parameters of the policy at iteration $t$, $\delta$ is the learning rate, and the policy gradient is
\begin{equation}
\label{gradient}
\begin{aligned}
\nabla_{\bm{\theta}} J(\bm{\theta})=&\nabla_{\bm{\theta}}\bar{A}(\bm{\theta})-\lambda \nabla_{\bm{\theta}}f_{\rm{KL}}\left({\bm{\pi}}_{{\bm{\theta}}^{*}}(\bm{s},\bm{a}),{\bm{\pi}}_{{\bm{\theta}}}(\bm{s},\bm{a})\right),\\
\approx & \frac{1}{K}\sum\limits_{k = 1}^K {{R(\bm{a}^*_k|\bm{s})}\frac{\nabla_{\bm{\theta}}{\bm{\pi}}_{{\bm{\theta}}}(\bm{s},\bm{a}^*_k)}{{\bm{\pi}}_{{\bm{\theta}}^*}(\bm{s},\bm{a}^*_k)}}\\
&+\lambda \sum\limits_{k = 1}^K{{\bm{\pi}}_{{\bm{\theta}}^*}(\bm{s},\bm{a}^*_k)}\nabla_{\bm{\theta}}\log\frac{{\bm{\pi}}_{{\bm{\theta}}}(\bm{s},\bm{a}^*_k)}{{\bm{\pi}}_{{\bm{\theta}}^*}(\bm{s},\bm{a}^*_k)},\\
=&\frac{1}{K}\sum\limits_{k = 1}^K {\frac{{\bm{\pi}}_{{\bm{\theta}}}(\bm{s},\bm{a}^*_k)}{{\bm{\pi}}_{{\bm{\theta}}^*}(\bm{s},\bm{a}^*_k)}{R(\bm{a}^*_k|\bm{s})}\frac{\nabla_{\bm{\theta}}{\bm{\pi}}_{{\bm{\theta}}}(\bm{s},\bm{a}^*_k)}{{\bm{\pi}}_{{\bm{\theta}}}(\bm{s},\bm{a}^*_k)}}\\
&+\lambda \sum\limits_{k = 1}^K\nabla_{\bm{\theta}}\log{{\bm{\pi}}_{{\bm{\theta}}}(\bm{s},\bm{a}^*_k)},\\
=&\frac{1}{K}\sum\limits_{k = 1}^K {\frac{{\bm{\pi}}_{{\bm{\theta}}}(\bm{s},\bm{a}^*_k)}{{\bm{\pi}}_{{\bm{\theta}}^*}(\bm{s},\bm{a}^*_k)}{R(\bm{a}^*_k|\bm{s})}\nabla_{\bm{\theta}}\log{{\bm{\pi}}_{{\bm{\theta}}}(\bm{s},\bm{a}^*_k)}}\\
&+\lambda \sum\limits_{k = 1}^K\nabla_{\bm{\theta}}\log{{\bm{\pi}}_{{\bm{\theta}}}(\bm{s},\bm{a}^*_k)}.\\
\end{aligned}
\end{equation}
From (\ref{SGA}) and (\ref{gradient}), we can see that the learning rate can be adjusted by the penalty coefficient $\lambda$ to guarantee the convergence of the APPO algorithm.
Hence, after $T$ iterations based on (\ref{SGA}), we update the penalty coefficient $\lambda$ as
\begin{equation}
\label{lambda}
\lambda \leftarrow \left\{ {\begin{array}{*{20}{l}}
{\eta\lambda,\;f_{\rm{KL}}\left({\bm{\pi}}_{{\bm{\theta}}^*}(\bm{s},\bm{a}),{\bm{\pi}}_{{\bm{\theta}}^{(T)}}(\bm{s},\bm{a})\right)>{\mu}_{\rm{high}},}\\
{\;\frac{\lambda}{\eta}\;,\;f_{\rm{KL}}\left({\bm{\pi}}_{{\bm{\theta}}^*}(\bm{s},\bm{a}),{\bm{\pi}}_{{\bm{\theta}}^{(T)}}(\bm{s},\bm{a})\right) < {\mu}_{\rm{low}}},\\
{\;\lambda\;,\; \rm{otherwise}},
\end{array}} \right.
\end{equation} 
where ${\mu}_{\rm{high}}=1+{\tau}$ and ${\mu}_{\rm{low}}=1-{\tau}$ are the thresholds that trigger the update of the penalty coefficient $\lambda$ and $\eta>1$ is a coefficient used to adjust the learning rate according to the difference between the stored policy ${\bm{\pi}}_{{\bm{\theta}}^*}(\bm{s},\bm{a})$ and the updated policy ${\bm{\pi}}_{{\bm{\theta}}^{(T)}}(\bm{s},\bm{a})$.
Then, the BS replace the stored policy as 
\begin{equation}
\label{replace}
{\bm{\theta}}^* = {\bm{\theta}}^{(T)},
\end{equation}
and resamples the actions for next $T$ policy update iterations.

By iteratively updating the policy until the proposed APPO algorithm converges, the policy parameter $\bm{\theta}$ can find the relation between the importance distributions of all semantic information and the total MSS.
Hence, the policy for RB allocation and semantic information selection that can achieve maximum total MSS of all recovered texts can be obtained \cite{VPG_huan}.
The specific training process of the proposed APPO algorithm is summarized in \textbf{Algorithm~1}.

\begin{algorithm}[t]
\small 
\caption{Training process of the proposed APPO algorithm.}
\begin{algorithmic}[1]
\STATE \textbf{Input:} Text $L_i$ required to transmit to each user,  delay threshold $D$, and interference $I_q$ of each RB.
\STATE \textbf{Initialize:} Parameters ${\bm{\theta}}^*$ generated randomly, semantic information extraction model, text recovery model, penalty coefficient $\lambda$, threshold $\tau$, coefficient $\eta$.
\STATE Obtain the importance distribution $\bm{f}({\cal{G}}_i)$ of each semantic information based on (\ref{distribution}).
\REPEAT
\STATE Store the policy ${\bm{\pi}}_{{\bm{\theta}}^*}(\bm{s},\bm{a})$ and collect $K$ trajectories ${\cal{K}}=\{\bm{a}_1,\ldots,\bm{a}_K\}$ using ${\bm{\pi}}_{{\bm{\theta}}^*}(\bm{s},\bm{a})$.
\FOR {$i = 1 \to T$} 
\STATE Update the parameters of the policy ${\bm{\pi}}_{{\bm{\theta}}}(\bm{s},\bm{a})$ based on (\ref{SGA}).
\ENDFOR
\STATE Update the penalty coefficient $\lambda$ based on (\ref{lambda}).
\STATE Replace the stored policy based on (\ref{replace}).
\UNTIL the objective function defined in (\ref{loss}) converges.
\end{algorithmic}
\label{algorithm_1}
\end{algorithm} 

\vspace{-0.2cm}
\subsection{Complexity and Convergence of APPO}
Next, we analyze the computational complexity and the convergence of the proposed APPO algorithm for resource allocation and partial semantic information determination.
The complexity of the APPO algorithm lies in calculating the importance distribution by using the attention network and determining the resource allocation by using the trained policy.
We first analyze the complexity of the attention network-based semantic importance evaluator.
From (\ref{correlation}), the complexity of calculating the correlation between semantic triple $\bm{\varepsilon}_i^g$ and token $w_{i,n}$ in original text $L_i$ is ${\cal{O}}\left(B_i^g(2D_aD_x+{D_a}^2)\right)$, where ${D_a} \times {D_x}$ is the size of parameter matrices $\bm{W}_{{\rm{tri}}}$ and $\bm{W}_{{\rm{tok}}}$ of the attention network.
We assume that $D_a \ll D_x$ (e.g., $D_a=64$ and $D_x=500$ as done in \cite{graph_transformer}).
Hence, we have ${\cal{O}}\left(B_i^g(2D_aD_x+{D_a}^2)\right)={\cal{O}}\left(B_i^g{D_a}{D_x}\right)$.
From (\ref{importance}) and (\ref{distribution}), the complexity of using the attention network to calculate the importance distribution for all semantic information can be given as ${\cal{O}}\left(\sum\limits_{i = 1}^U {\sum\limits_{g = 1}^{{G_i}} {{N_i}{B_i^g}{D_a}{D_x}}}\right)$.
Next, we investigate the complexity for resource allocation that depends on the size of the policy parameter $\bm{\theta}$.
The size of the policy parameter $\bm{\theta}$ depends on the size of action space $\cal{A}$ and the size of state space $\cal{S}$.
The action space $\cal{A}$ is a set of all possible resource allocation whose size is ${|\cal{A}|}=\frac{{U!}}{{(U - Q)!}}$.
Based on (\ref{distribution}), the size of state space $\cal{S}$ is ${|\cal{S}|}=\sum\limits_{i = 1}^U {{G_i}}$.
Then, the complexity of using the trained policy to determine the resource allocation is ${\cal{O}}\left(\frac{{U!}}{{(U - Q)!}}\left(\sum\limits_{i = 1}^U {{G_i}}\right)\prod\limits_{l = 2}^{L - 1} {{H_l}}\right)$, where $L$ is the number of layers in deep neural network used to train the policy and ${H_l}$ is the number of the neurons in layer $l$.
In consequence, the computational complexity of the proposed APPO algorithm for resource allocation and semantic information selection will be 
\begin{equation}
\label{complexity}	
{\cal{O}}\left(\sum\limits_{i = 1}^U {\sum\limits_{g = 1}^{{G_i}} {{N_i}{B_i^g}{D_a}{D_x}}}\!+\!\frac{{U!}}{{(U \!-\!Q)!}}\!\left(\sum\limits_{i = 1}^U {{G_i}}\right)\!\prod\limits_{l = 2}^{L - 1} {{H_l}}\right).
\end{equation}
From (\ref{complexity}), given the structure of the proposed APPO model (i.e., $D_a$, $D_x$, and $H_l$), the complexity of APPO depends on the number of users, the number of RBs, and the number of triples in each extracted semantic information.

With regards to the convergence of the proposed APPO algorithm, we can directly use the result of \cite[Theorem 1]{TRPO} which shows that the policy of an APPO algorithm will converge to a locally optimal solution when it satisfies the following conditions. 

\itshape \text{{\bf{Lemma 1}} (follows from \cite{TRPO})}{\rm{:}} \upshape
 The proposed APPO algorithm used to solve the total MSS maximization problem (\ref{optimization}) converges if the following conditions are satisfied:
\begin{enumerate}
\item[i)] $J(\bm{\theta})$ defined in (\ref{loss}) is a lower bound function of the objective function $\bar{A}({\bm{\theta}})$ defined in (\ref{expected_reward}) that the BS aims to optimize.
\item[ii)] The actions sampled according to a stored policy ${\bm{\pi}}_{{\bm{\theta}}^*}(\bm{s},\bm{a})$ can be used to evaluate and iteratively update the policy ${\bm{\pi}}_{{\bm{\theta}}}(\bm{s},\bm{a})$.
\item[iii)] The value of the objective function $\bar{A}({\bm{\theta}})$ monotonically increases as the policy ${\bm{\theta}}$ is iteratively updated based on (\ref{SGA}).
\end{enumerate}

\itshape \text{Proof:}  \upshape
Next, we prove that our proposed algorithm satisfies the conditions from \cite{TRPO}.
For condition i), the design of the lower bound function $J(\bm{\theta})$ follows \cite{TRPO} and, hence, the proof in \cite{TRPO} is still hold.
For condition ii), the difference between the stored policy ${\bm{\pi}}_{{\bm{\theta}}^*}(\bm{s},\bm{a})$ and the policy ${\bm{\pi}}_{{\bm{\theta}}}(\bm{s},\bm{a})$ to be updated is controlled by the penalty of the Kullback Leibler divergence as we specified in (\ref{loss}).
For condition iii), the learning rate of the APPO algorithm can be adjusted based on (\ref{lambda}) so that the value of the objective function $\bar{A}({\bm{\theta}})$ can monotonically increase according to the proof in \cite{TRPO}.
\hfill $\Box$

\begin{table*}\footnotesize
\newcommand{\tabincell}[2]{\begin{tabular}{@{}#1@{}}#1.6\end{tabular}}
\renewcommand\arraystretch{1}
\caption[table]{System parameters \cite{VR_TWC_chenmingzhe_huan, PPO, BERT}}
\centering
\begin{tabular}{|c|c|c|c|c|c|}
\hline
\textbf{Parameters}&\textbf{Description}&\textbf{Value}\\
\hline
$Q $ & Number of RBs & 10 \\
\hline
$W$ & Bandwidth of each RB &  2 MHz\\
\hline
$P$ & Transmit power of the BS &1 W \\
\hline
 $N_0$ & Noise power spectral density & -174 dBm/Hz \\
\hline
$D$ & Transmission delay threshold & $0.1$ ms \\
\hline
 $O$&Data size of each token& 80 bits  \\
\hline
 $D_a$ &Dimension of hidden layer in attention network & 64 \\
\hline
 $\varphi$ & Weight parameter in (11) &0.5 \\
 \hline
  $L$ &Number of layers in DNN used to output policies & 3\\
\hline
 $K$ &Batch size of policy update& 100\\
 \hline
$\eta$ & Coefficient used to adjust the learning rate &2\\
\hline
$D_x$ & Dimension of token vector &500 \\
\hline
 $\tau$ &Threshold of triggering the penalty coefficient &0.8\\
 \hline
 \end{tabular}
\end{table*}

From Lemma 1, we see that, the convergence of the proposed APPO algorithm depends on the lower bound function $J(\bm{\theta})$ and the dynamic learning rate. 
In particular, as the policy is iteratively updated, the lower bound function $J(\bm{\theta})$ locally approximates the objective function $\bar{A}({\bm{\theta}})$ and the learning rate decreases until the APPO algorithm converges to the locally optimal solution of problem (\ref{optimization}).
According to \cite{TRPO}, the proposed APPO algorithm for solving (\ref{optimization}) will converge.

\vspace{-0.2cm}
\section{Simulation Results and Analysis}
\label{sec:4}
In our simulations, a circular network is considered with one BS and $U=30$ uniformly distributed users. Other parameters are listed in Table \uppercase\expandafter{\romannumeral2}. 
We use the semantic information extraction model in \cite{IE_system} and the text recovery model in \cite{graph_transformer}.
The text datasets used to train the proposed APPO algorithm are the abstract generation dataset (AGENDA)\cite{AGENDA} which consists of 40 thousand paper titles and abstracts from the proceedings of 12 top artificial intelligence conferences and the DocRED dataset \cite{DocRED} which consists of 5053 Wikipedia documents.
For comparison purposes, we consider five baselines: 
baseline a) being the APG algorithm proposed in \cite{wyy2021GC},  
baseline b) being a deep Q network integrated with attention network (ADQN) algorithm,
baseline c) that optimizes RB allocation using the proposed RL solution and directly transmits the original text data, 
baseline d) that optimizes RB allocation using the proposed RL solution and randomly selects semantic triples to be transmitted,
and baseline e) being the semantic framework proposed in \cite{Deep_Semantic}.
Here, for baseline c), the BS transmits the original texts to the users token by token until the delay threshold is reached \cite{SCN}. 
The MSS measures the semantic similarity between the original text and the partial text received by each user.

\begin{figure*}[t]
\centering
\setlength{\belowcaptionskip}{0cm}
\includegraphics[width=0.88\linewidth]{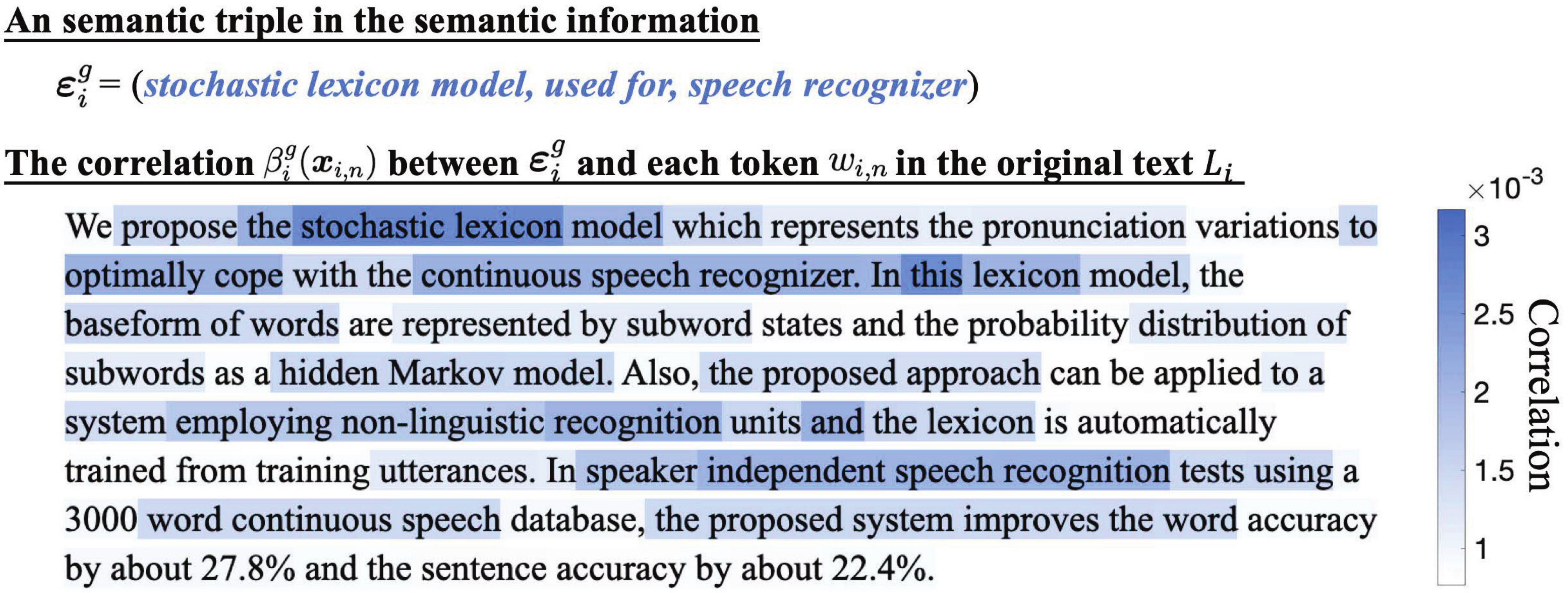}
\caption{The correlation between an example of the semantic triple and the original text.}
\label{fig3}
\end{figure*}

Figure~\ref{fig3} shows the correlation (as defined in (\ref{correlation})) between a semantic triple and the tokens in the original text.
In Fig.~\ref{fig3}, the importance of the triple ``(\emph{stochastic lexicon model}, \emph{used for}, \emph{speech recognize})" is shown as the sum of correlations (i.e., (\ref{importance})).
In particular, we use different colors to represent the correlation between the semantic triple (``\emph{stochastic lexicon model}, \emph{used for}, \emph{speech recognize}") and different tokens in the original text.
As the correlations between the semantic triple and the tokens in the original text increase, the color used to mark the tokens changes from white to blue.
From Fig.~\ref{fig3}, we can see that the expression ``\emph{the proposed approach}" in the fourth line of the original text is highly correlated with the example triple ``(\emph{stochastic lexicon model}, \emph{used for}, \emph{speech recognizer})".
This implies that the attention networks can extract correlations between the triples and the original text according to the meaning of the tokens in the context.
This is due to the fact that, compared with the traditional approaches that obtain the correlation between two token sequences by word frequency statistics, the attention networks can generate the representation for each token according to the context of the token and calculate the correlations based on these representations.
In consequence, the attention networks enable the BS to obtain the importance of the triples in the semantic information.

\begin{figure*}[t]
\centering
\setlength{\belowcaptionskip}{-0.4cm}
\includegraphics[width=0.88\linewidth]{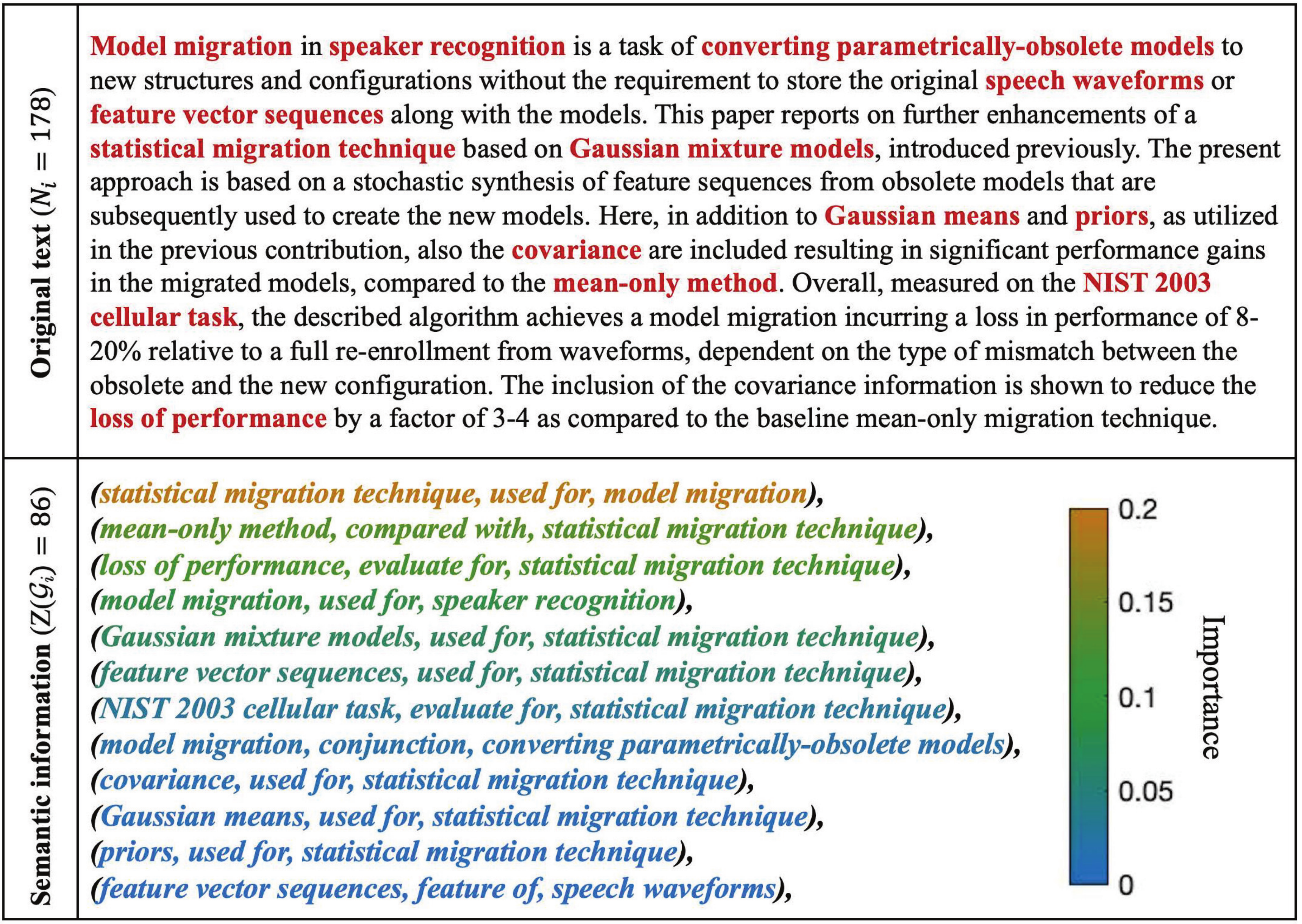}
\caption{An example of the original text and the extracted semantic information.}
\label{fig8}
\end{figure*}

Figure~\ref{fig8} shows an example of the original text, the extracted semantic information ${\cal{G}}_i$, and the importance distribution of the semantic triples in ${\cal{G}}_i$.
In Fig.~\ref{fig8}, the entities recognized in the original text are shown in red and the extracted semantic triples are written in different colors according to their importance.
In particular, as the importance of a triple increases, the color of that semantic triple changes from blue to yellow.
For example, the importance of the yellow triple (i.e., ``\emph{statistical migration technique, used for, model migration}") is 0.1923 while the importance of the blue triple (i.e., ``\emph{feature vector sequences, feature of, speech waveforms}") is 0.0385.
The sum of the importance of all the triples in a semantic information is 1.
From Fig.~\ref{fig8}, we can see that, using our proposed semantic communication framework, the BS needs to transmit only
86 tokens of the semantic information instead of 178 tokens of the original text data to the user.
Therefore, in this example, the proposed framework can reduce the size of the data that needs to be transmitted by 51.7\%.

\begin{figure*}
\centering 
\setlength{\abovecaptionskip}{0cm}
\setlength{\belowcaptionskip}{-0.5cm}
\subfigbottomskip=7pt 
\subfigcapskip=0pt 
\subfigure[The solution of the proposed APPO algorithm.]{
\label{fig9a}
\includegraphics[width=0.88\linewidth]{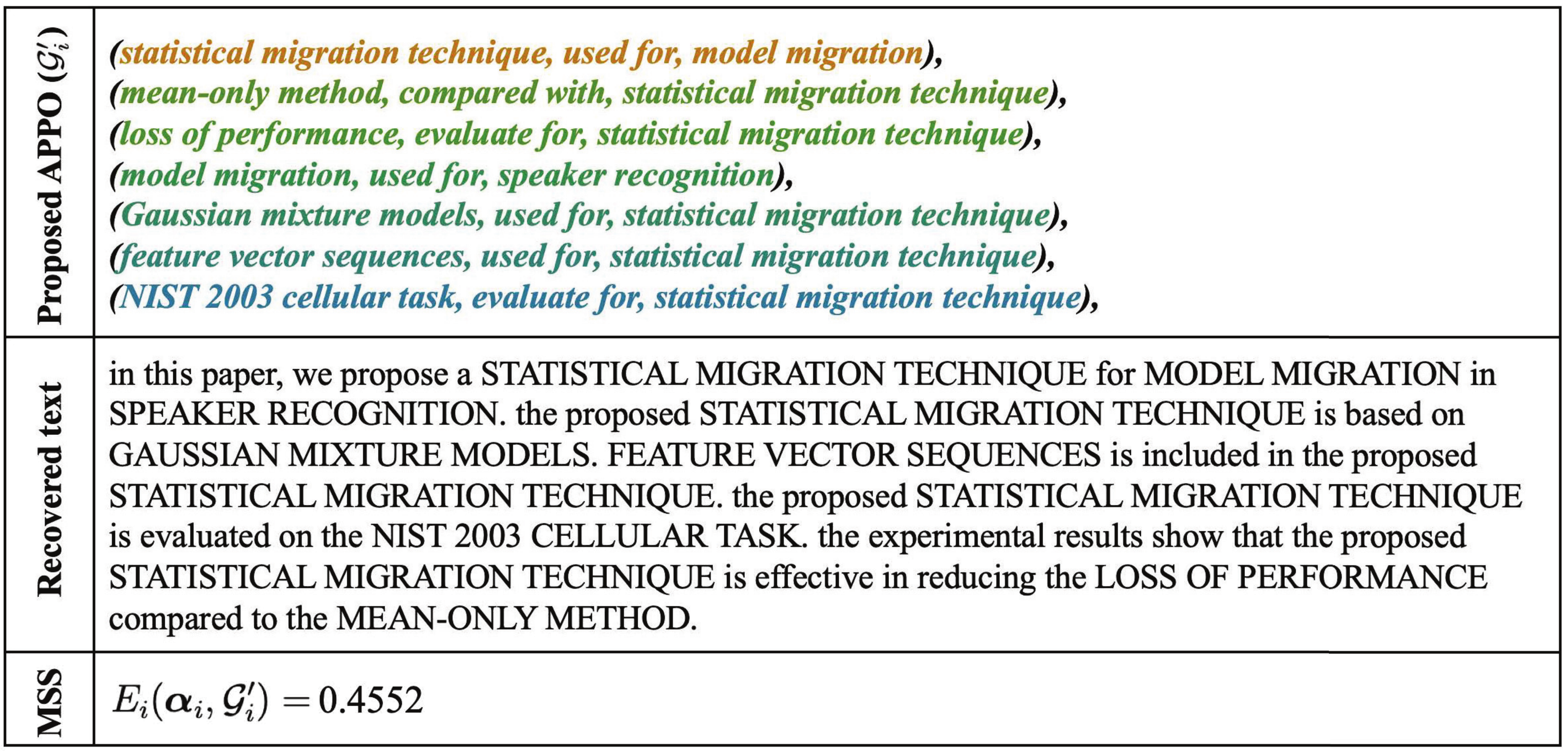}}\\
\subfigure[The solution of the baseline a).]{
\label{fig9b}
\includegraphics[width=0.88\linewidth]{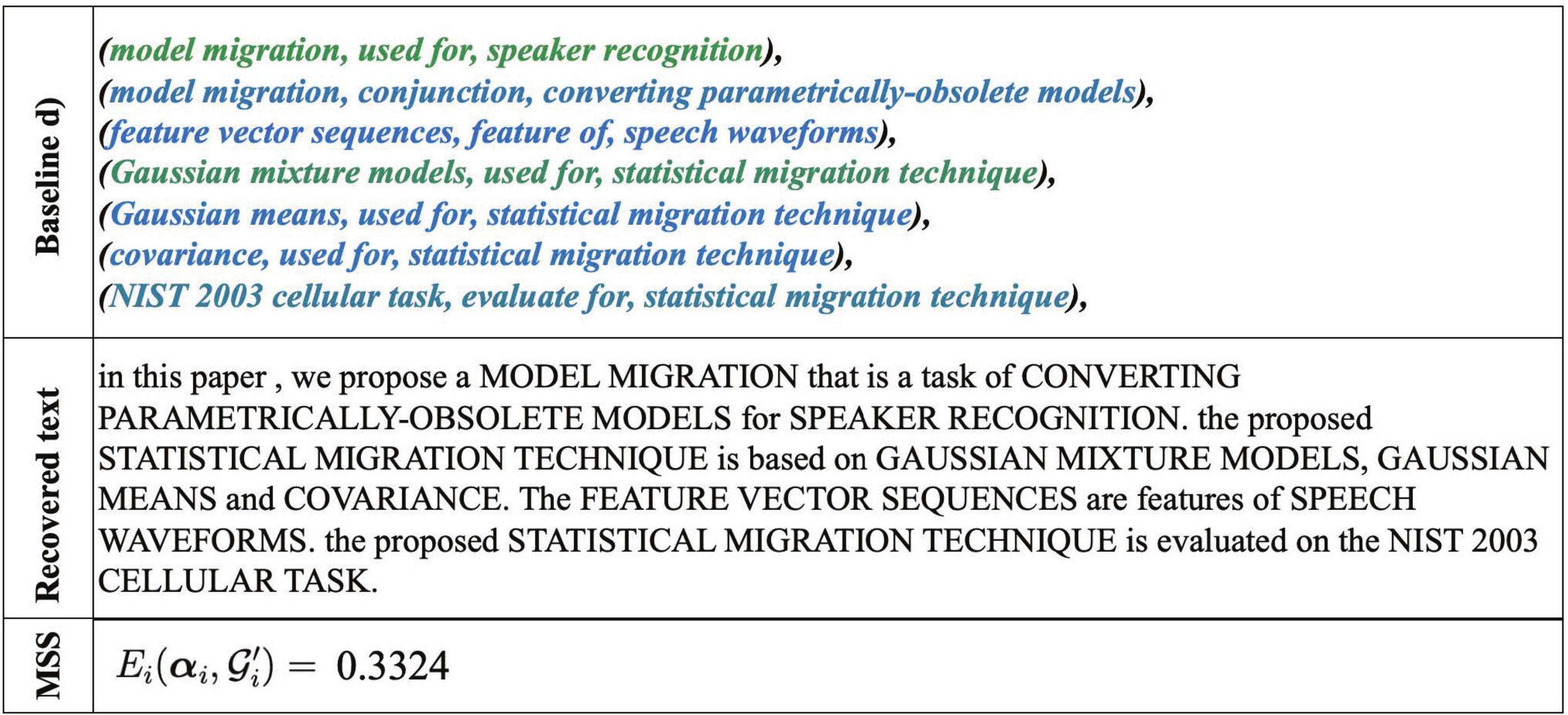}}\\
\subfigure[The solution of the baseline b).]{
\label{fig9c}
\includegraphics[width=0.88\linewidth]{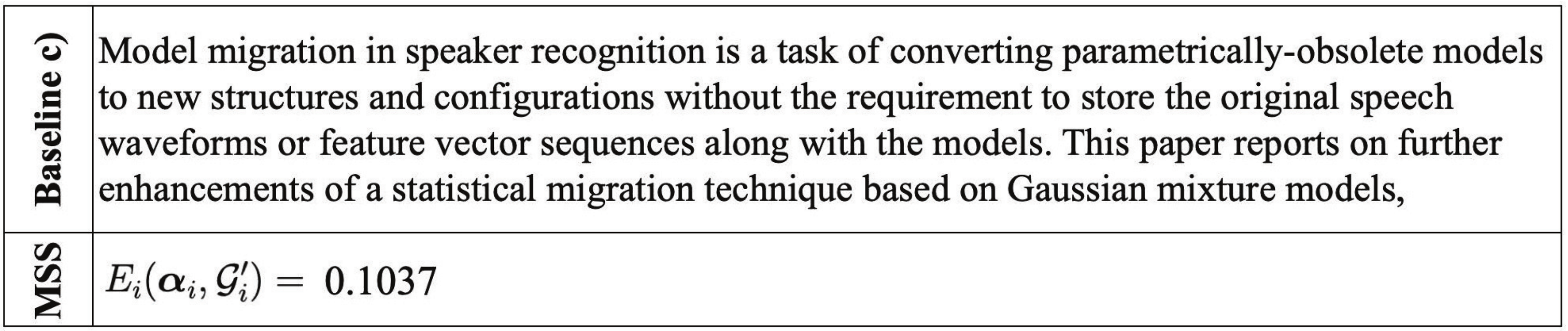}}
\caption{An example of the transmitted data and the corresponding recovered text.}
\label{fig9}
\end{figure*}

Figure~\ref{fig9} shows the partial semantic information that will be transmitted and the corresponding recovered text for the example in Fig.~\ref{fig8}.
Fig.~\ref{fig9a} shows the partial semantic information ${\cal{G}}'_i$ selected by the proposed APPO algorithm and the text recovered using ${\cal{G}}'_i$.
In Fig.~\ref{fig9a}, the BS uses the APPO algorithm to select the most important triples in the extracted semantic information and transmits them to the user.
Figs.~\ref{fig9b} and \ref{fig9c} show the results of baselines a) and b).
From Figs.~\ref{fig9a} and \ref{fig9b}, we see that the proposed algorithm can improve 36.7\% total MSS compared to baseline a).
This is because the proposed APPO algorithm can evaluate the importance of the triples in the semantic information and, then determines the RB allocation and partial semantic information to be transmitted.
Therefore, the text recovered using the most important semantic triples contains the main meaning of the original text.
From Figs.~\ref{fig9a} and \ref{fig9c}, we observe that the proposed algorithm can improve 2-fold total MSS compared to baseline b). 
This gain stems from the fact that the proposed semantic communication framework can extract the meaning of the original data and transmit it to each user. 

\begin{figure}[t]
\centering
\setlength{\belowcaptionskip}{-0.5cm}
\includegraphics[width=8.5cm]{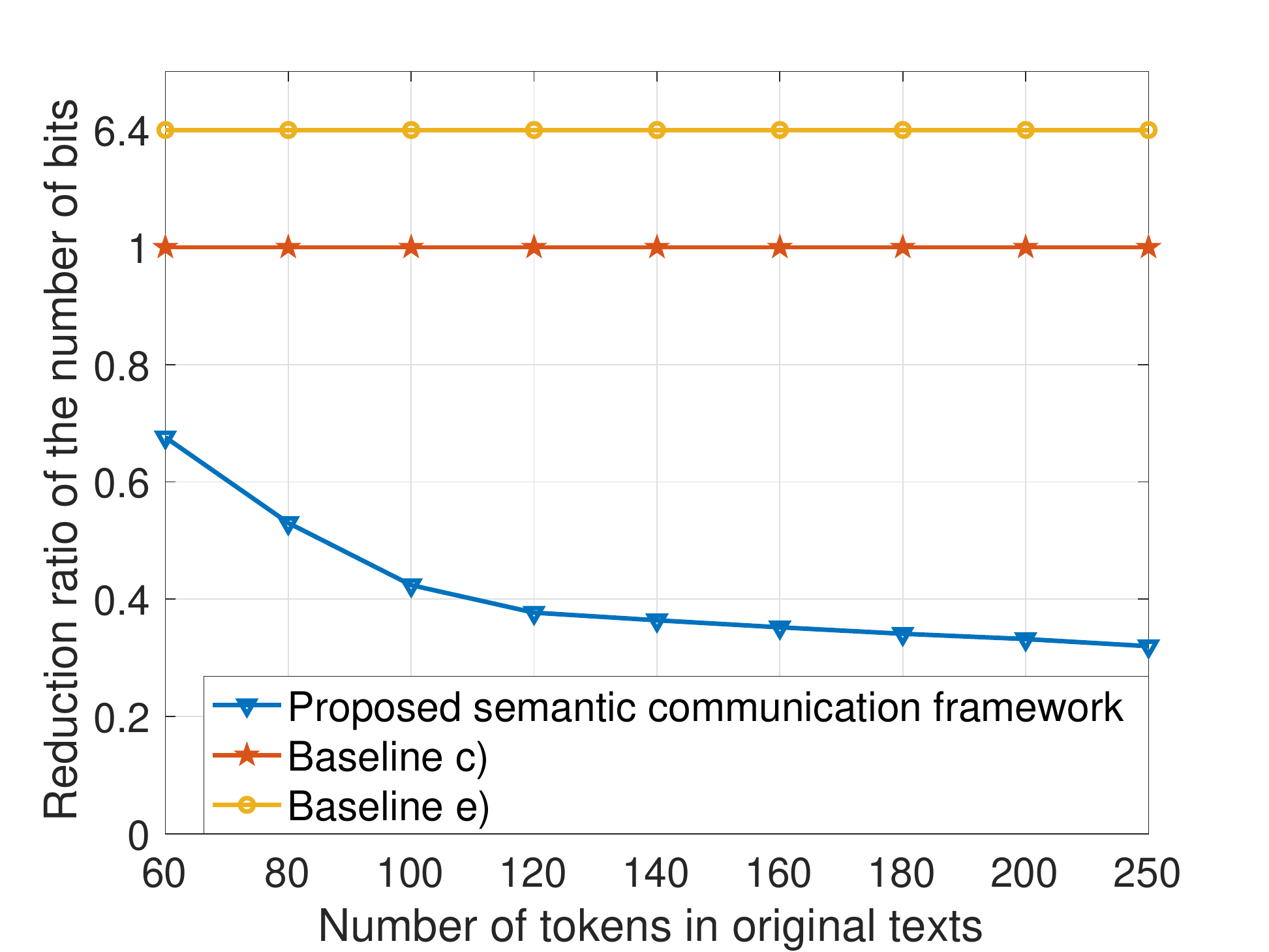}
\caption{The reduction ratio of the data size using the semantic communication framework.}
\label{fig4}
\end{figure}

Figure~\ref{fig4} shows how the ratio of the data size of the semantic information to the data size of the original text changes as the number of tokens in the original text changes.
We assume that the data size of an english letter is 8 bits and each token consists of 10 letters.
Hence, the number of bits used to encode a token is 80 (i.e., in (\ref{data_size}), $O=80$) \cite{standard}.
Baseline e) uses 16-dimensional feature vectors extracted by the deep neural networks to represent the semantic information of the tokens in the original text \cite{Deep_Semantic}.
We also assume that a decimal consists of 32 bits.
From Fig.~\ref{fig4} we can see that, compare to baseline c) and baseline e), the proposed semantic communication framework can reduce the size of the data that needs to be transmitted by up to 41.3\% and 84\%, respectively
The reason is that our proposed framework enables the BS to extract the meaning of the texts and model the extracted meaning by a knowledge graph.
From Fig.~\ref{fig4}, we also observe that, as the number of tokens in the original texts increases, the ratio of the size of the semantic information to the size of the original text decreases.
This is due to the fact that the semantic information in multiple sentences in long texts can be expressed by a single semantic triple.
Hence, extracting the semantic information from long texts can significantly reduce the number of tokens that need to be transmitted.

\begin{figure*}[t]
\centering
\setlength{\belowcaptionskip}{-0.5cm}
\includegraphics[width=0.88\linewidth]{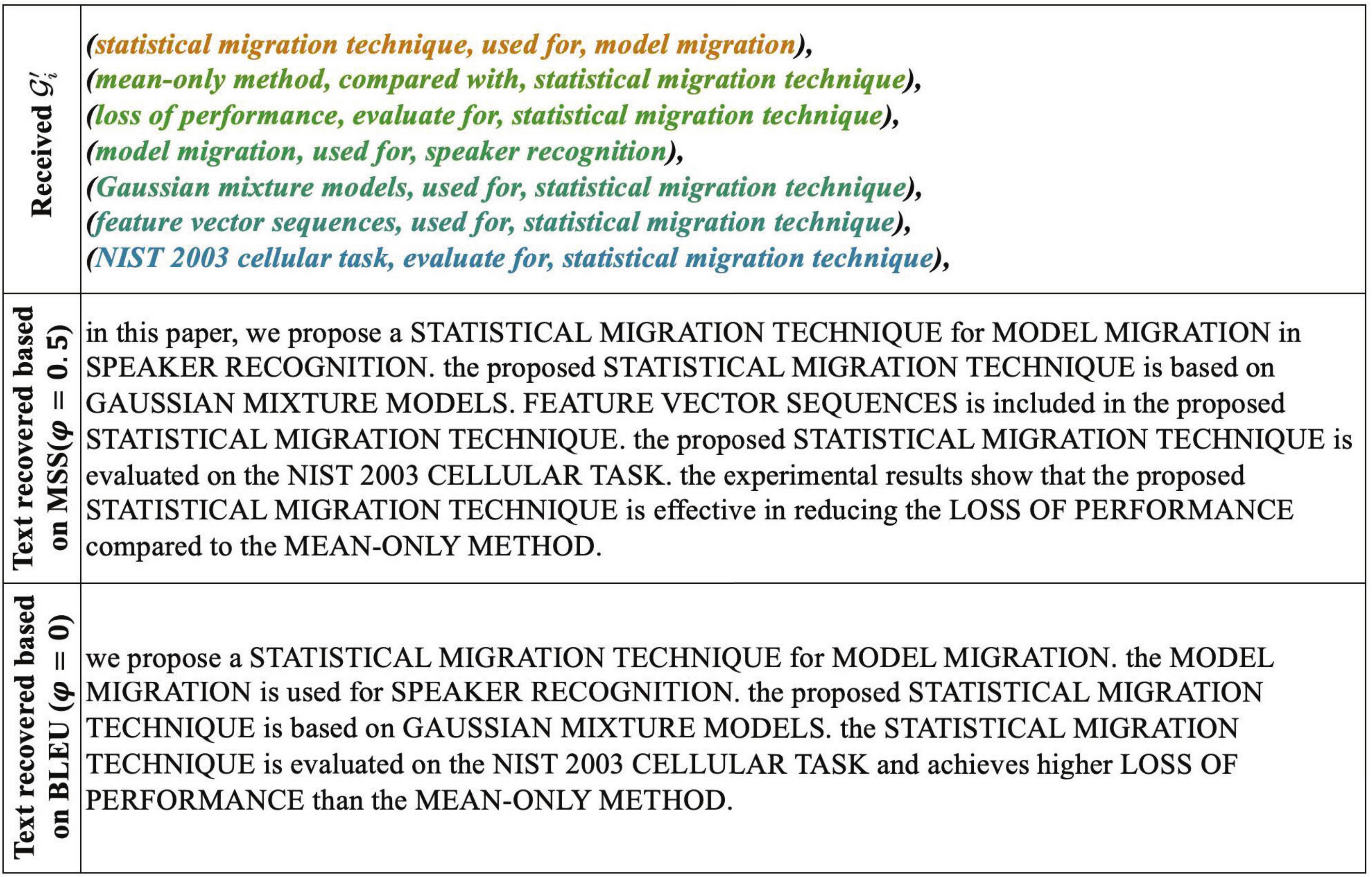}
\caption{The text recovered based on different metric.}
\label{fig7}
\end{figure*}

Figure~\ref{fig7} shows examples of the texts recovered by the graph-to-text generation models that are trained via the loss function different metrics.
In particular, we include the result of the graph-to-text generation model that aims to maximize the proposed MSS and BLEU in \cite{BLEU}.
From Fig.~\ref{fig7}, we can see that the text recovered using the MSS-based text generation model contains all semantic triples in the received semantic information and all semantic triples is arranged with articulated textual context.
In contrast, the text recovered using the BLEU-based text generation model misses the semantic triple ``\emph{feature vector sequences, used for, statistical migration technique}".
This is due to the fact that, compared to the BLEU metric that only captures the accuracy of the recovered text, the proposed MSS can capture both accuracy and completeness of the recovered text.
From Fig.~\ref{fig7}, we can also see that, compared to the BLEU-based text generation model that copies semantic triples from the received semantic information for text recovery, the MSS-based text generation model can generate additional descriptions based on the context of the semantic triples.
For example, the MSS-based text generation model generates ``\emph{the experimental results show that...}".
This is because an additional penalty for short text is considered in the definition of MSS.

\begin{figure}[t]
\centering
\setlength{\belowcaptionskip}{-0.5cm}
\includegraphics[width=8cm]{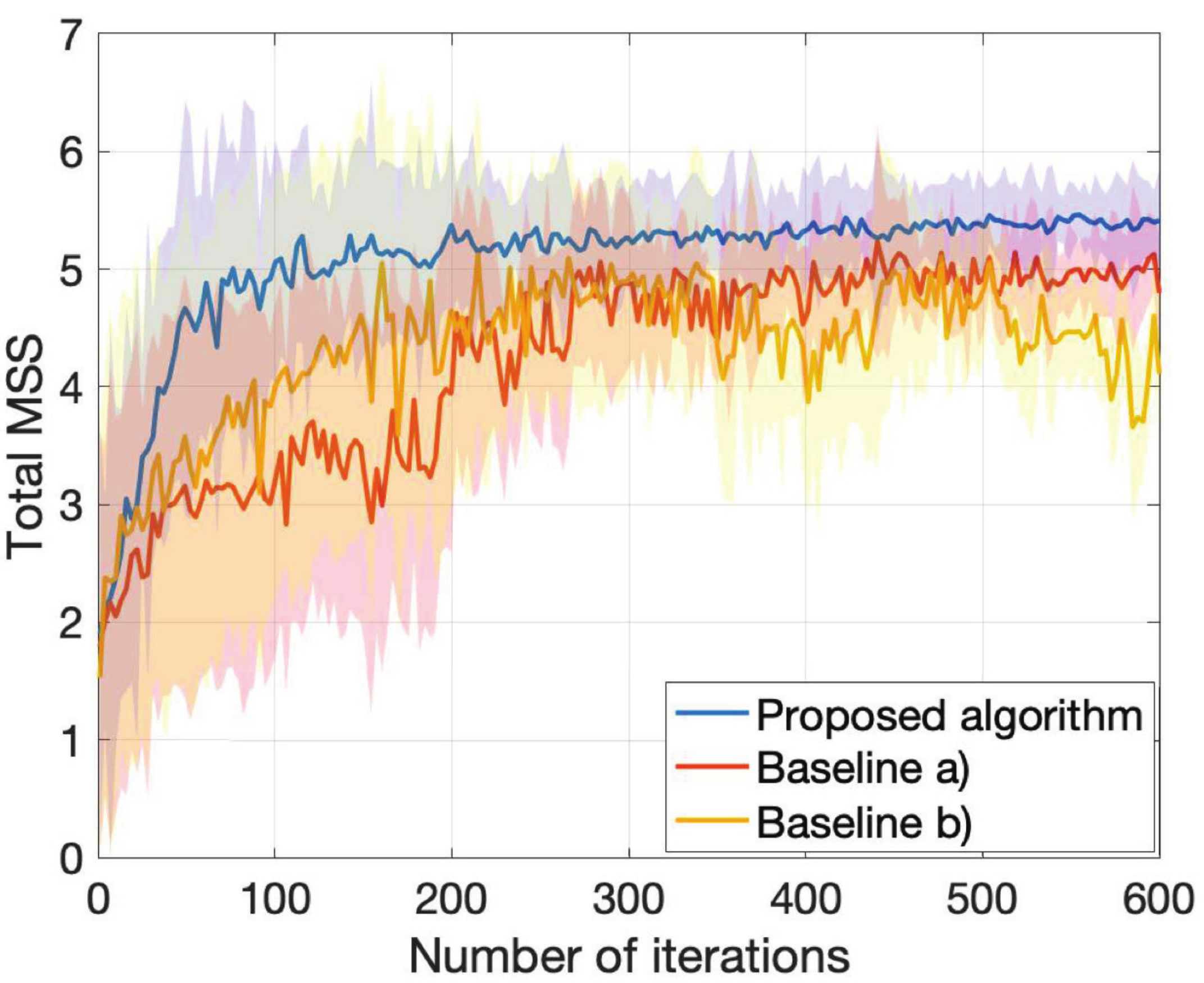}
\caption{The training process of the considered algorithms.}
\label{fig5}
\end{figure}

Figure~\ref{fig5} shows the convergences of the APPO, APG, and DQN integrated with attention networks (ADQN) algorithms in a semantic communication-enabled network with 10 RBs.
The line and shadow are the mean and standard deviation computed over 10 independent runs.
From Fig.~\ref{fig5}, we observe that, compared to the APG and ADQN algorithms that require about 300 iterations and 250 iterations to reach convergence, respectively, the proposed APPO algorithm converges after 100 iterations.
This stems from the fact that the proposed APPO algorithm can adjust the learning rate at each iteration so as to speed up the convergence.
The improvement in the convergence speed indicates that the proposed APPO algorithm can reduce the time and energy overhead of the training process in practical semantic communication systems.
Figure~\ref{fig5} also shows that, the proposed APPO algorithm can improve the MSS by 11.2\% and 18.5\%, respectively, compared with the APG algorithm and ADQN algorithm.
The reason is that, when the updated policy approximates the locally optimal solution, the APPO algorithm can further improve the total MSS by decreasing the learning rate.
In practice, this indicates that the APPO algorithm enables the BS to effectively allocate the limited RBs according to the importance distribution of semantic information so as to improve the performance of the proposed semantic communication framework.
In Fig.~\ref{fig5}, we can also see that, the APPO algorithm achieves up to 53.3\% and 60.1\% reduction in the standard deviation of the total MSS compared to the APG algorithm.
In practice, this implies that, the performance of the proposed APPO algorithm is more stable than that of the APG and ADQN algorithms.

\begin{figure}[t]
\centering
\setlength{\belowcaptionskip}{-0.5cm}
\includegraphics[width=8cm]{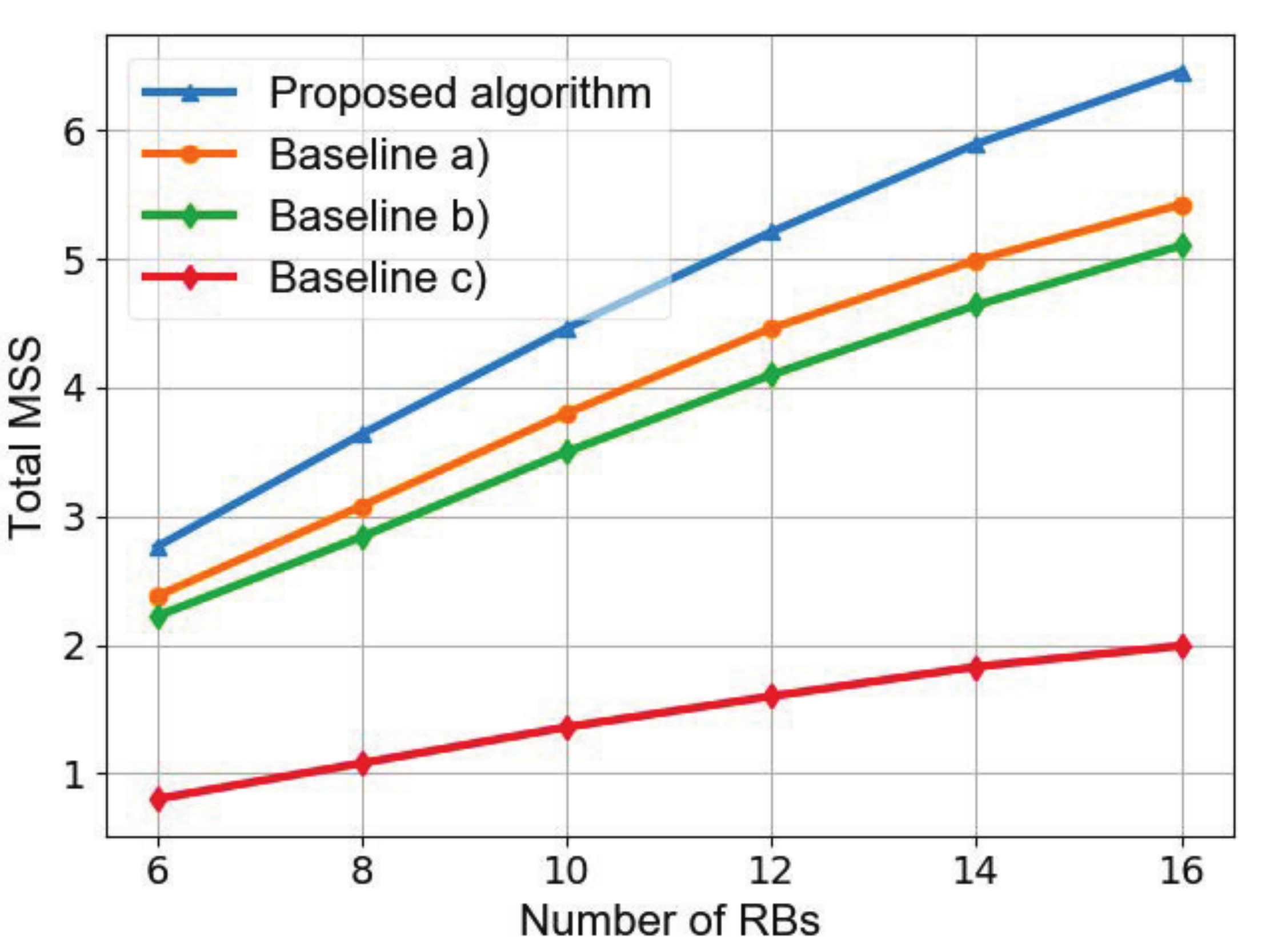}
\caption{The total MSS as the number of RBs varies.}
\label{fig6}
\end{figure}

Figure~\ref{fig6} shows how the total MSS changes as the number of RBs varies. 
This figure is simulated using the DocRED dataset.
From Fig.~\ref{fig6}, we can see that the proposed APPO algorithm can improve 17.8\%, 26.8\% and 2-fold total MSS on DocRED dataset compared to baselines a), b) and c), respectively.
The 17.8\% and 26.8\% gains stem from the fact that the proposed APPO algorithm dynamically adjusts the learning rate during training process.
Hence, the APPO algorithm is guaranteed to obtain a locally optimal RB allocation policy for partial semantic information transmission.
The 2-fold\% gain stems from the fact that the proposed algorithm enables the BS to transmit the most important semantic triples using the limited wireless resource thus significantly improving the transmission efficiency.

\begin{figure}[t]
\centering
\setlength{\belowcaptionskip}{-0.5cm}
\includegraphics[width=8cm]{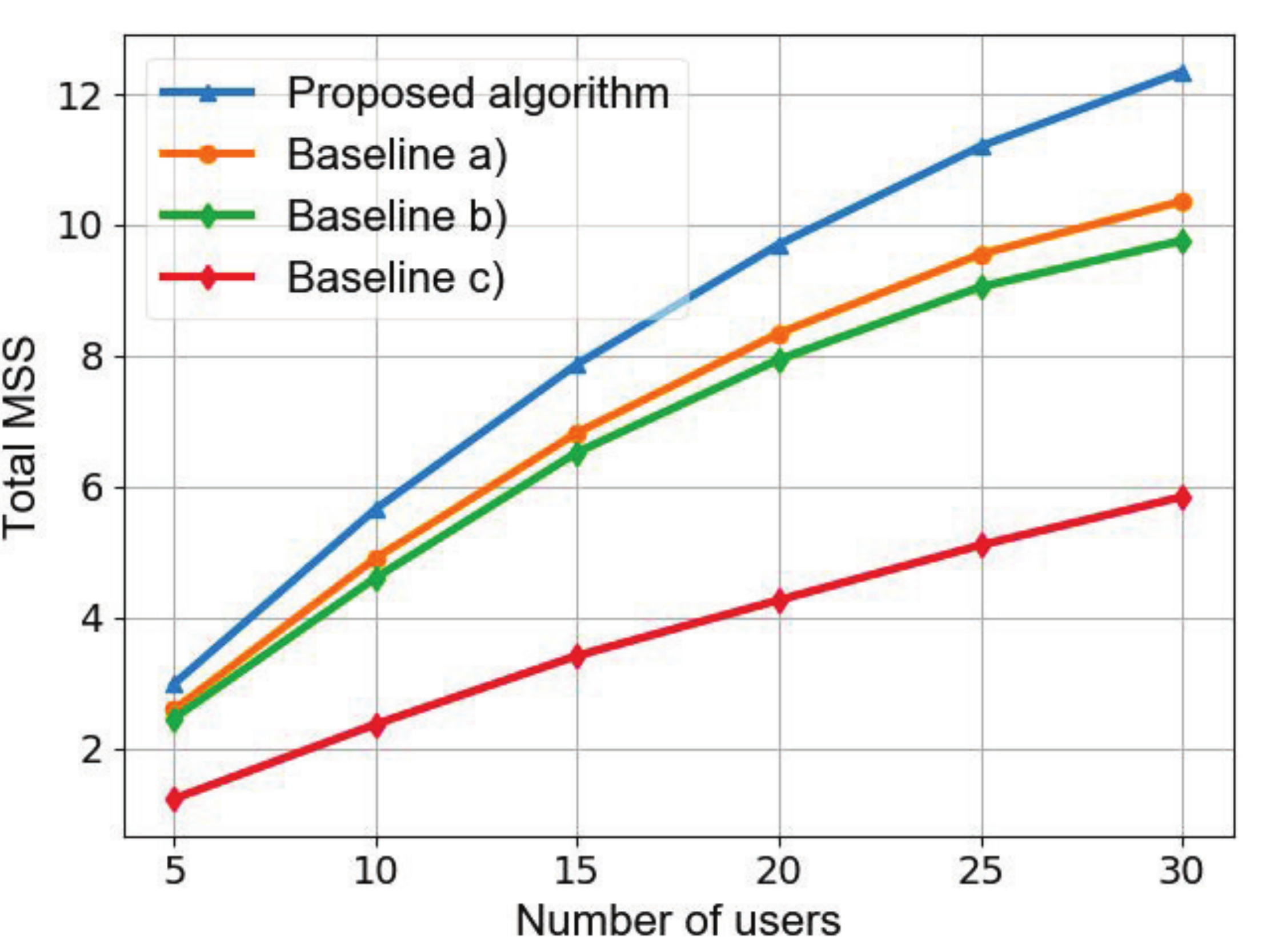}
\caption{The total MSS as the number of users varies.}
\label{user_num}
\end{figure}

Figure~\ref{user_num} shows how the total MSS changes as the number of users varies. 
This figure is simulated using the AGENDA dataset.
From Fig.~\ref{user_num}, we can see that the proposed algorithm can serve 30 users.
From Fig.~\ref{user_num}, we can also see that, the proposed APPO algorithm can improve 16.9\%, 22.3\% and 1-fold total MSS compared to baselines a), b) and c), respectively.
The 16.9\% and 22.3\% gains stem from the fact that the proposed APPO algorithm can build the relationship between the importance distribution of the semantic information and the total MSS, thus effectively finding the locally optimal RB allocation policy for partial semantic information transmission.
The 1-fold\% gain stems from the fact that the proposed algorithm uses the knowledge graph to model the semantic information and transmit the important semantic triples thus significantly improving the transmission efficiency.

\begin{figure}[t]
\centering
\setlength{\belowcaptionskip}{-0.5cm}
\includegraphics[width=9cm]{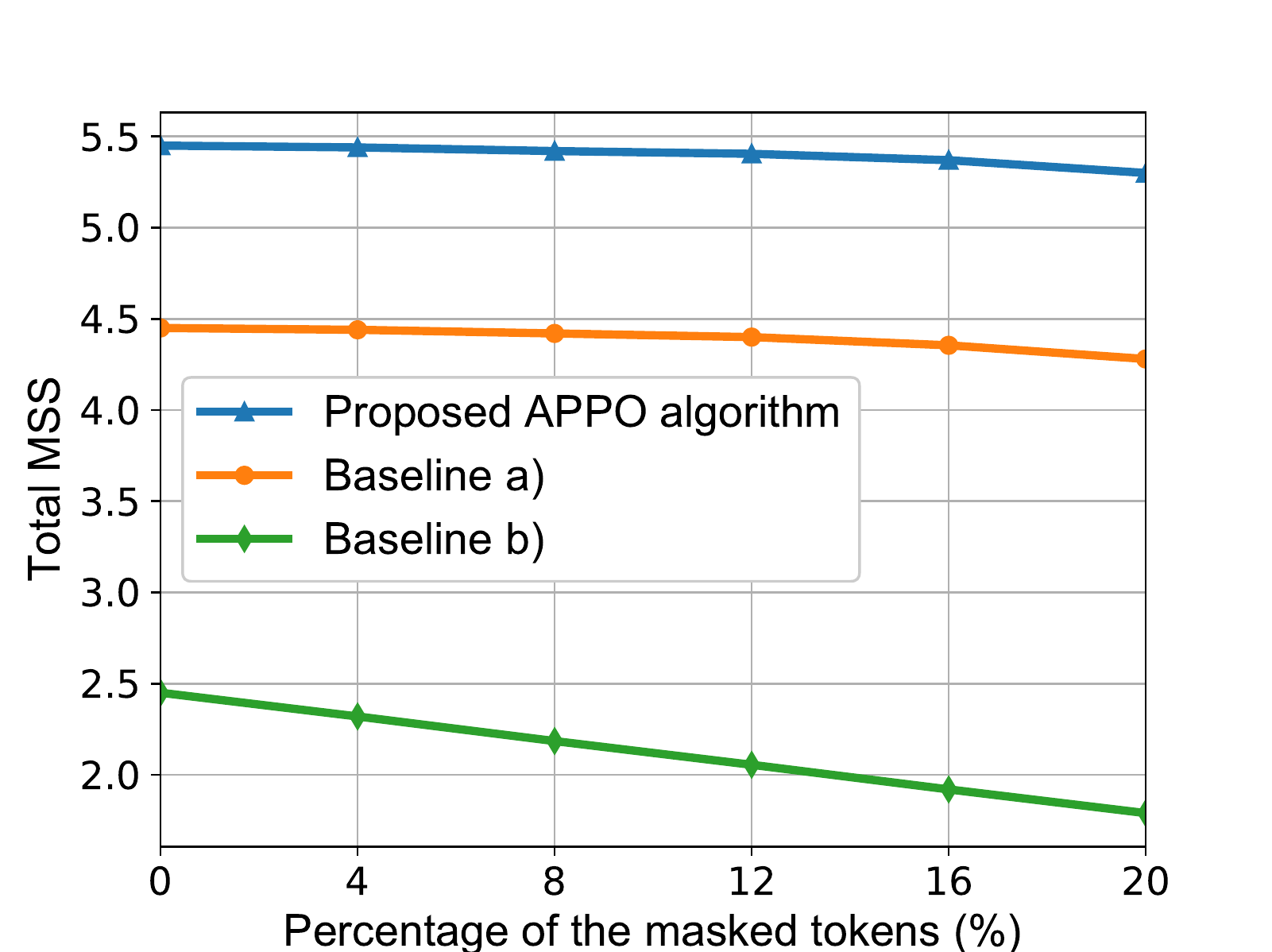}
\caption{The total MSS as the percentage of the masked tokens
varies.}
\label{robusts}
\end{figure} 

Figure~\ref{robusts} shows how the total MSS changes as the percentage of the masked tokens varies.
In Fig.~\ref{robusts}, we randomly mask a subset of the tokens in the original texts in the AGENDA dataset to verify the robustness of the proposed semantic communication framework.
The obtained texts with masks are used to simulate texts containing typos in actual scenarios.
From Fig~\ref{robusts}, we can see that, as the percentage of masked tokens increases, the total MSS of the proposed APPO algorithm, baseline a), and baseline b) decreases by 2.75\%, 3.8\%, and 67\%, respectively.
The reason why the total MSS of the proposed APPO algorithm and baseline a) remain stable is because the occasional masks do not change the semantic information of the original texts.
The semantic information extraction and original text recovery approaches enable the proposed semantic communication framework to ignore the masks in original texts.
The 67\% MSS reduction of baseline b) stems from the fact that the standard communication networks directly transmit the texts with masks to each user.

\vspace{-0.3cm}
\section{Conclusion}
\label{sec:5}
In this paper, we have proposed a semantic driven wireless networks.
We have modeled the semantic information of the textual data by a KG.
To measure the performance of the semantic communications, we have introduced a new metric, MSS, that captures the semantic accuracy and semantic completeness between the original text and the recovered text.
We have jointly considered the wireless resource limitations, transmission delay requirements, and the performance of the semantic communications and formulated an optimization problem whose goal is to maximize the total MSS by optimizing the RB allocation for partial semantic information transmission.
To solve this problem, we have developed an APPO algorithm that can obtain the importance distribution of the triples in the semantic information and then build the relationship between the importance distribution and the total MSS.
Hence, the proposed APPO algorithm enables the BS to find the policies for RB allocation and semantic information selection for maximizing the total MSS.
Compared with the traditional RL algorithms, the proposed APPO dynamically adjusts the learning rate during training process, thus improving the total MSS, the convergence speed, and the stability.
Simulation results have demonstrated that the proposed semantic communication framework can significantly reduce the size of data required to transmit and increase the total MSS.






%



\bibliographystyle{IEEEbib}
\def\baselinestretch{1}
\bibliography{Semantic}

\begin{thebibliography}{10}

\bibitem{wyy2021GC}
Y.~{Wang}, M.~{Chen}, W.~{Saad}, T.~{Luo}, S.~{Cui}, and H.~V. {Poor},
\newblock ``Performance optimization for semantic communications: An
  attention-based learning approach,''
\newblock in {\em Proc. IEEE Global Communications Conference}, Madrid, Spain,
  Dec. 2021.

\bibitem{6G_walid}
W.~{Saad}, M.~{Bennis}, and M.~{Chen},
\newblock ``A vision of 6{G} wireless systems: Applications, trends,
  technologies, and open research problems,''
\newblock {\em IEEE Network}, vol. 34, no. 3, pp. 134--142, May 2020.

\bibitem{Semantic_6G}
E.~C. {Strinati} and S.~{Barbarossa},
\newblock ``{6G} networks: Beyond shannon towards semantic and goal-oriented
  communications,''
\newblock {\em Computer Networks}, vol. 190, pp. 107930, May 2021.

\bibitem{Deep_Semantic}
H.~{Xie}, Z.~{Qin}, G.~{Li}, and B.~{Juang},
\newblock ``Deep learning enabled semantic communication systems,''
\newblock {\em IEEE Transactions on Signal Processing}, vol. 69, pp.
  2663--2675, April 2021.

\bibitem{Semantic_Magizine}
G.~{Shi}, Y.~{Xiao}, Y.~{Li}, and X.~{Xie},
\newblock ``From semantic communication to semantic-aware networking: Model,
  architecture, and open problems,''
\newblock {\em IEEE Communications Magazine}, vol. 59, no. 8, pp. 44--50, Aug.
  2021.

\bibitem{Semantic_IOT}
H.~{Xie} and Z.~{Qin},
\newblock ``A lite distributed semantic communication system for internet of
  things,''
\newblock {\em IEEE Journal on Selected Areas in Communications}, vol. 39, no.
  1, pp. 142--153, Jan. 2021.

\bibitem{Semantic_Significance}
E.~{Uysal}, O.~{Kaya}, A.~{Ephremides}, J.~{Gross}, M.~{Codreanu},
  P.~{Popovski}, M.~{Assaad}, G.~{Liva}, A.~{Munari}, T.~{Soleymani},
  B.~{Soret}, and K.~H. {Johansson},
\newblock ``Semantic communications in networked systems,''
\newblock 2021,
\newblock Available: \url{https://arxiv.org/abs/2103.05391}.

\bibitem{Semantic_AoI1}
M.~{Kountouris} and N.~{Pappas},
\newblock ``Semantics-empowered communication for networked intelligent
  systems,''
\newblock {\em IEEE Communications Magazine}, vol. 59, no. 6, pp. 96--102, June
  2021.

\bibitem{Semantic_Hierarchical}
G.~{Shi}, D.~{Gao}, X.~{Song}, J.~{Chai}, M.~{Yang}, X.~{Xie}, L.~{Li}, and
  X.~{Li},
\newblock ``A new communication paradigm: {From} bit accuracy to semantic
  fidelity,''
\newblock 2021,
\newblock Available: \url{https://arxiv.org/abs/2101.12649}.

\bibitem{NSW_2011_theory}
J.~{Bao}, P.~{Basu}, M.~{Dean}, C.~{Partridge}, A.~{Swami}, W.~{Leland}, and
  J.~A. {Hendler},
\newblock ``Towards a theory of semantic communication,''
\newblock in {\em Proc. IEEE Network Science Workshop}, West Point, NY, USA,
  June 2011.

\bibitem{Semantic_AoI2}
A.~{Maatouk}, M.~{Assaad}, and A.~{Ephremides},
\newblock ``The age of incorrect information: an enabler of semantics-empowered
  communication,''
\newblock 2020,
\newblock Available: \url{https://arxiv.org/abs/2012.13214}.

\bibitem{Semantic_Speech}
Z.~{Weng} and Z.~{Qin},
\newblock ``Semantic communication systems for speech transmission,''
\newblock {\em IEEE Journal on Selected Areas in Communications}, vol. 39, no.
  8, pp. 2434--2444, Aug. 2021.

\bibitem{Semantic_Game}
B.~{Güler}, A.~{Yener}, and A.~{Swami},
\newblock ``The semantic communication game,''
\newblock {\em IEEE Transactions on Cognitive Communications and Networking},
  vol. 4, no. 4, pp. 787--802, Dec. 2018.

\bibitem{resource}
O.~{Semiari}, W.~{Saad}, M.~{Bennis}, and M.~{Debbah},
\newblock ``Integrated millimeter wave and {Sub-6} {GHz} wireless networks: A
  roadmap for joint mobile broadband and ultra-reliable low-latency
  communications,''
\newblock {\em IEEE Wireless Communications}, vol. 26, no. 2, pp. 109--115,
  April 2019.

\bibitem{VR_TWC_chenmingzhe_huan}
M.~{Chen}, Z.~{Yang}, W.~{Saad}, C.~{Yin}, H.~V. {Poor}, and S.~{Cui},
\newblock ``A joint learning and communications framework for federated
  learning over wireless networks,''
\newblock {\em IEEE Transactions on Wireless Communications}, vol. 20, no. 1,
  pp. 269--283, Jan. 2021.

\bibitem{resource_zhanghaijun}
Y.~{Li}, H.~{Zhang}, and K.~{Long},
\newblock ``Joint resource, trajectory, and artificial noise optimization in
  secure driven {3D} {UAVs} with {NOMA} and imperfect {CSI},''
\newblock {\em IEEE Journal on Selected Areas in Communications}, vol. 39, no.
  11, pp. 3363--3377, Nov. 2021.

\bibitem{JSTSP_WSH}
S.~{Wang}, M.~{Chen}, Z.~{Yang}, C.~{Yin}, W.~{Saad}, S.~{Cui}, and H.~V.
  {Poor},
\newblock ``Distributed reinforcement learning for age of information
  minimization in real-time iot systems,''
\newblock {\em IEEE Journal of Selected Topics in Signal Processing}, to
  appear, 2022.

\bibitem{resource_liuyuanwei}
W.~{Ahsan}, W.~{Yi}, Z.~{Qin}, Y.~{Liu}, and A.{Nallanathan},
\newblock ``Resource allocation in uplink {NOMA-IoT} networks: A
  reinforcement-learning approach,''
\newblock {\em IEEE Transactions on Wireless Communications}, Aug. 2021.

\bibitem{resource_xiongkangdusit}
Z.~{Xiong}, J.~{Kang}, D.~{Niyato}, P.~{Wang}, and H.~V. {Poor},
\newblock ``Cloud/edge computing service management in blockchain networks:
  Multi-leader multi-follower game-based {ADMM} for pricing,''
\newblock {\em IEEE Transactions on Services Computing}, vol. 13, no. 2, pp.
  356--367, March-April 2020.

\bibitem{goal_Walid}
M.~{Karimzadeh}, W.~{Saad}, and M.~{Debbah},
\newblock ``Common language for goal-oriented semantic communications: A
  curriculum learning framework,''
\newblock in {\em Proc. IEEE International Conference on Communications},
  Seoul, South Korea, May 2022.

\bibitem{speech1}
D.~{Michelsanti}, Z.~{Tan}, S.~{Zhang}, Y.~{Xu}, M.~{Yu}, D.~{Yu}, and
  J.~{Jensen},
\newblock ``An overview of deep-learning-based audio-visual speech enhancement
  and separation,''
\newblock {\em IEEE/ACM Transactions on Audio, Speech, and Language
  Processing}, vol. 29, pp. 1368--1396, March 2021.

\bibitem{image1}
F.~{Liu}, E.~{Bugliarello}, M.~E. {Ponti}, S.~{Reddy}, N.~{Collier}, and
  D.~{Elliott},
\newblock ``Visually grounded reasoning across languages and cultures,''
\newblock in {\em Proc. Conference on Empirical Methods in Natural Language
  Processing}, Punta Cana, Dominican Republic, Nov. 2021, pp. 10467--10485.

\bibitem{KG}
S.~{Ji}, S.~{Pan}, E.~{Cambria}, P.~{Marttinen}, and P.~S. Yu,
\newblock ``A survey on knowledge graphs: Representation, acquisition, and
  applications,''
\newblock {\em IEEE Transactions on Neural Networks and Learning Systems}, to
  appear, 2021.

\bibitem{IE_system}
Y.~{Luan}, L.~{He}, M.~{Ostendorf}, and H.~{Hajishirzi},
\newblock ``Multi-task identification of entities, relations, and coreference
  for scientific knowledge graph construction,''
\newblock in {\em Proc. Conference on Empirical Methods in Natural Language
  Processing}, Brussels, Belgium, Oct. 2018.

\bibitem{CNN_relation}
Y.~{Lin}, S.~{Shen}, Z.~{Liu}, H.~{Luan}, and M.~{Sun},
\newblock ``Neural relation extraction with selective attention over
  instances,''
\newblock in {\em Proc. Annual Meeting of the Association for Computational
  Linguistics}, Berlin, Germany, Aug. 2016, pp. 2124--2133.

\bibitem{chen_2021}
M.~{Chen}, N.~{Shlezinger}, H.~V. {Poor}, Y.~C. {Eldar}, and S.~{Cui},
\newblock ``Communication efficient federated learning,''
\newblock {\em Proc. National Academy of Sciences of the United States of
  America}, vol. 118, no. 17, Apr. 2021.

\bibitem{graph_transformer}
R.~{Koncel-Kedziorski}, D.~{Bekal}, Y.~{Luan}, M.~{Lapata}, and
  H.~{Hajishirzi},
\newblock ``Text generation from knowledge graphs with graph transformers,''
\newblock in {\em Proc. Conference of the North {A}merican Chapter of the
  Association for Computational Linguistics}, Minneapolis, Minnesota, June
  2019.

\bibitem{BLEU}
K.~{Papineni}, S.~{Roukos}, T.~{Ward}, and W.~{Zhu},
\newblock ``{BLEU}: a method for automatic evaluation of machine translation,''
\newblock in {\em Proc. Annual Meeting of the Association for Computational
  Linguistics}, Philadelphia, PA, USA, July 2002, pp. 311--318.

\bibitem{METEOR}
S.~{Banerjee} and A.~{Lavie},
\newblock ``{METEOR}: An automatic metric for {MT} evaluation with improved
  correlation with human judgments,''
\newblock in {\em Proc. Annual Meeting of the Association for Computational
  Linguistics Workshop}, Ann Arbor, MI, USA, June 2005.

\bibitem{PPO}
P.~{Dhariwal} A. {Radford} O.~{Klimov} J.~{Schulman}, F.~{Wolski},
\newblock ``Proximal policy optimization algorithms,''
\newblock 2017,
\newblock Available: \url{https://arxiv.org/abs/1707.06347}.

\bibitem{attention}
A.~{Vaswani}, N.~{Shazeer}, N.~{Parmar}, J.~{Uszkoreit}, L.~{Jones}, A.~N.
  {Gomez}, U.~{Kaiser}, and I.~{Polosukhin},
\newblock ``Attention is all you need,''
\newblock in {\em Proc. International Conference on Neural Information
  Processing Systems}, Long Beach, CA, USA, Dec. 2017.

\bibitem{BERT}
J.~{Devlin}, M.~{Chang}, K.~{Lee}, and K.~{Toutanova},
\newblock ``{BERT}: Pre-training of deep bidirectional transformers for
  language understanding,''
\newblock 2018,
\newblock Available: \url{https://arxiv.org/abs/1810.04805}.

\bibitem{TRPO}
J.~{Schulman}, S.~{Levine}, P.{Moritz}, M.~{Jordan}, and P.~{Abbeel},
\newblock ``Trust region policy optimization,''
\newblock in {\em Proc. International Conference on Machine Learning}, Lille,
  UK, July 2015, pp. 1889--1897.

\bibitem{VPG_huan}
Y.~{Hu}, M.~{Chen}, W.~{Saad}, H.~V. {Poor}, and S.~{Cui},
\newblock ``Distributed multi-agent meta learning for trajectory design in
  wireless drone networks,''
\newblock {\em IEEE Journal on Selected Areas in Communications}, vol. 39, no.
  10, pp. 3177--3192, Oct. 2021.

\bibitem{AGENDA}
W.~{Ammar}, D.~{Groeneveld}, C.~{Bhagavatula}, L.~{Beltagy}, M.~{Crawford},
  D.~{Downey}, J.~{Dunkelberger}, A.~{Elgohary}, S.~{Feldman}, V.~A. {Ha},
  R.~M. {Kinney}, S.~{Kohlmeier}, K.~{Lo}, T.~C. {Murray}, H.~{Ooi}, M.~E.
  {Peters}, J.~L. {Power}, S.~{Skjonsberg}, L.~{Wang}, C.~{Wilhelm}, Z.~{Yuan},
  M.~{Zuylen}, and O.~{Etzioni},
\newblock ``Construction of the literature graph in semantic scholar,''
\newblock 2018,
\newblock Available: \url{https://arxiv.org/abs/1805.02262}.

\bibitem{DocRED}
Y.~{Yao}, D.~{Ye}, P.~{Li}, X.~{Han}, Y.~{Lin}, Z.~{Liu}, Z.~{Liu}, L.~{Huang},
  J.~{Zhou}, and M.~{Sun},
\newblock ``{DocRED}: A large-scale document-level relation extraction
  dataset,''
\newblock in {\em Proc. Association for Computational Linguistics}, Florence,
  Italy, July 2019.

\bibitem{SCN}
C.~{Liu}, C.~{Guo}, Y.~{Yang}, and N.~{Jiang},
\newblock ``Adaptable semantic compression and resource allocation for
  task-oriented communications,''
\newblock 2022,
\newblock Available: \url{https://arxiv.org/abs/2204.08910}.

\bibitem{standard}
International~Organization for Standardization and
  International~Electrotechnical Commission,
\newblock {\em Information technology - {ISO} 8-bit code for information
  interchange - Structure and rules for implementation},
\newblock {ISO/IEC} 4873:1991, 1991.

\end{thebibliography}

\end{document}